\newif\ifUseOnlineSupplements  
\UseOnlineSupplementsfalse
\documentclass[11pt,letterpaper]{article}
 
\RequirePackage{amsmath,amsfonts,amssymb,ifthen,url,graphicx,color,array,amsthm}
\usepackage[margin=1in]{geometry}
\usepackage{natbib}  

\newtheorem{theorem}{Theorem} 
\newtheorem{definition}{Definition}
\newtheorem{proposition}{Proposition}
\newtheorem{lemma}{Lemma} 
\newtheorem{corollary}{Corollary}

\def\argmin{\mathop{\rm arg\,min}}%

\usepackage[toc,page]{appendix} 

\newenvironment{fiamline}{}{}  
\usepackage[framemethod=TikZ]{mdframed}
\def\Halmos{\hfill \(\square \) }% \square <--> \blackbox (?)
\def\proof#1{\begin{fiamline} \textit{#1}}
\def\endproof{ \end{fiamline} } 
\usepackage{floatrow}  
\usepackage[labelfont={bf,it},textfont=bf]{caption}

\usepackage[linesnumbered,ruled,vlined,resetcount]{algorithm2e}
\SetKwFor{ForEach}{for each}{do}{end for each}
\SetKwInput{KwData}{Input}
\SetKwInput{KwResult}{Output}
\let\oldnl\nl% Store \nl in \oldnl
\newcommand{\nonl}{\renewcommand{\nl}{\let\nl\oldnl}}% Remove line number for one line
\usepackage{csquotes}
\usepackage{anyfontsize}
\usepackage[hidelinks]{hyperref}
\usepackage[nameinlink]{cleveref}
\usepackage{tikz,xcolor}
\usepackage[outline]{contour}
\contourlength{2pt}
\usetikzlibrary{patterns}
\tikzset{
	pattern size/.store in=\mcSize, 
	pattern size = 5pt,
	pattern thickness/.store in=\mcThickness, 
	pattern thickness = 0.3pt,
	pattern radius/.store in=\mcRadius, 
	pattern radius = 1pt}
\makeatletter
\pgfutil@ifundefined{pgf@pattern@name@linedownleftright}{
	\pgfdeclarepatternformonly[\mcThickness,\mcSize]{linedownleftright}
	{\pgfqpoint{0pt}{-\mcThickness}}
	{\pgfpoint{\mcSize}{\mcSize}}
	{\pgfpoint{\mcSize}{\mcSize}}
	{
		\pgfsetcolor{\tikz@pattern@color}
		\pgfsetlinewidth{\mcThickness}
		\pgfpathmoveto{\pgfqpoint{0pt}{\mcSize}}
		\pgfpathlineto{\pgfpoint{\mcSize+\mcThickness}{-\mcThickness}}
		\pgfusepath{stroke}
}}
\makeatother

\tikzset{
	pattern size/.store in=\mcSize, 
	pattern size = 5pt,
	pattern thickness/.store in=\mcThickness, 
	pattern thickness = 0.3pt,
	pattern radius/.store in=\mcRadius, 
	pattern radius = 1pt}
\makeatletter
\pgfutil@ifundefined{pgf@pattern@name@lineupleftright}{
	\pgfdeclarepatternformonly[\mcThickness,\mcSize]{lineupleftright}
	{\pgfqpoint{0pt}{0pt}}
	{\pgfpoint{\mcSize+\mcThickness}{\mcSize+\mcThickness}}
	{\pgfpoint{\mcSize}{\mcSize}}
	{
		\pgfsetcolor{\tikz@pattern@color}
		\pgfsetlinewidth{\mcThickness}
		\pgfpathmoveto{\pgfqpoint{0pt}{0pt}}
		\pgfpathlineto{\pgfpoint{\mcSize+\mcThickness}{\mcSize+\mcThickness}}
		\pgfusepath{stroke}
}}
\makeatother

\tikzset{
	pattern size/.store in=\mcSize, 
	pattern size = 5pt,
	pattern thickness/.store in=\mcThickness, 
	pattern thickness = 0.3pt,
	pattern radius/.store in=\mcRadius, 
	pattern radius = 1pt}
\makeatletter
\pgfutil@ifundefined{pgf@pattern@name@linevertical}{
	\pgfdeclarepatternformonly[\mcThickness,\mcSize]{linevertical}
	{\pgfqpoint{-\mcThickness}{-\mcThickness}}
	{\pgfpoint{\mcSize}{\mcSize}}
	{\pgfpoint{\mcSize}{\mcSize}}
	{
		\pgfsetcolor{\tikz@pattern@color}
		\pgfsetlinewidth{\mcThickness}
		\pgfpathmoveto{\pgfpointorigin}
		\pgfpathlineto{\pgfpoint{0}{\mcSize}}
		\pgfusepath{stroke}
}}
\makeatother

\tikzset{every picture/.style={line width=0.75pt}} %set default line width to 0.75pt 
\usepackage{enumerate}
\usepackage{booktabs,multirow,tabularx}
\newcolumntype{C}{>{\centering\arraybackslash}X}
\newenvironment{heuristic}[1][]
{% Update algorithm name
  \let\c@algocf\c@hh% Update algorithm counter
  \begin{algorithm}[#1]%
    }{\end{algorithm}}
\crefname{heuristic}{heuristic}{heuristics}
\Crefname{heuristic}{Heuristic}{Heuristics}

\title{Bi-Criteria Multiple Knapsack Problem with Grouped Items}
\author{Francisco Castillo-Zunino, Pinar Keskinocak\\
H. Milton Stewart School of Industrial and Systems Engineering,\\
Georgia Institute of Technology\\
Atlanta, Georgia 30332\\
fj.castillo.zunino@gatech.edu,  pinar@isye.gatech.edu} 
\date{\today}
 
\begin{document}	 
\maketitle
% chktex-file 1 % cmd terminated with space
The multiple knapsack problem with grouped items aims to maximize rewards by assigning groups of items among multiple knapsacks, considering knapsack capacities.
Either all items in a group are assigned or none at all.
We propose algorithms which guarantee that rewards are not less than the optimal solution, with a bound on exceeded knapsack capacities.
To obtain capacity-feasible solutions, we propose a binary-search heuristic combined with these algorithms.
We test the performance of the algorithms and heuristics in an extensive set of experiments on randomly generated instances
and show they are efficient and effective, i.e., they run reasonably fast and generate good quality solutions.

% remove some chktex warnings
% chktex-file 1 % cmd terminated with space
% chktex-file 9 % ) expected, found } 
% chktex-file 10 % Solo `]' found.
% chktex-file 21 % "this command might not be intended" \left\{ bug

\section{Introduction}\label{SEC:intro}
In this paper we study the \textit{Multiple Knapsack Problem with Grouped Items} (GMKP), a generalization of the \textit{Multiple Knapsack Problem} (MKP).
MKP assigns a subset of items to multiple knapsacks, aiming to maximize the total reward without exceeding the capacity of any knapsack.
In GMKP, items are partitioned into groups, each group has a reward (if assigned), and either all or none of the items from a group are assigned to knapsacks (not necessarily to the same knapsack).

One motivation for GMKP comes from patient scheduling \citep{kirthana}, where patients may require multiple therapy sessions during their treatment (e.g., for rehabilitation).
In this case, each knapsack corresponds to a day (capacities are the time slots available in each day), each group of items corresponds to the therapy sessions that need to be scheduled for a patient, and each item corresponds to a session (weights representing the session duration).
A patient is scheduled only if their treatment can be scheduled entirely; i.e., all therapy sessions must be scheduled, or the patient needs to wait and be scheduled in the future.

Knapsack problems have been well-studied in the literature \citep{wilbaut2008survey}.
The single \textit{0/1 Knapsack Problem} (KP), is \( \mathcal{NP}\)-hard but is solvable in pseudo-polynomial time by dynamic programming \citep{horowitz1974computing}.
Several  \textit{Polynomial Time Approximation Schemes} (PTAS) and \textit{Fully PTAS} (FPTAS) exist for KP \citep{kellerer1999new}.
MKP is not solvable in pseudo-polynomial time; it is strongly \( \mathcal{NP}\)-hard \citep{martello1990knapsack}.
There is a PTAS, but not an FPTAS for MKP\@; even for two knapsacks \citep{chekuri2005polynomial}.
There is no PTAS nor constant-ratio approximation algorithm for GMKP \citep{chen2018packing}.

Since there are no efficient approximation algorithms for GMKP, we relax capacity constraints to find good solutions (in terms of rewards from assignments), with a bound on how much the capacities are exceeded.
Hence, our focus is on the \textit{bi-criteria GMKP} (bi-GMKP), where the goal is to simultaneously maximize the total reward and minimize the maximum exceeded knapsack capacity.
The capacity-relaxed bi-GMKP is also motivated by patient scheduling, where additional time slots can be added to the schedule by utilizing overtime or temporary personnel.

MKP and other similar problems are frequently solved with  meta-heuristics such as genetic algorithms \citep{liu2015scalable, khuri1994zero}, tabu search \citep{woodcock2010hybrid}, and swarm intelligence algorithms \citep{krause2013survey,liu2014new}.
Other approaches include exact methods such as branch and bound \citep{posta2012exact,martello1981bound}, cutting plane \citep{avella2010computational, ferreira1996solving}, and column generation algorithms \citep{forrest2006column}.
There are several variants of MKP with additional constraints, such as assignment restrictions \citep{dawande2000approximation}, color constraints \citep{forrest2006column}, and other variants of the generalized assignment problem \citep{oncan2007survey}.

For a special case of GMKP, when all knapsack capacities are equal and the heaviest group weighs at most \(2/3\) of the total capacity, parameterized-approximation algorithms were proposed by~\cite{chen2018packing}.
~\cite{adany2016all} proposed a PTAS for a
generalized assignment problem with grouped items that has additional constraints:
there is a limit on the number of items per group,
and knapsacks can accommodate at most one item from each group.
The algorithms proposed by~\cite{chen2018packing} and~\cite{adany2016all} guarantee feasiblity while sacrificing rewards;
the algorithms proposed in this paper sacrifice feasibility (bounded by a maximum exceeded knapsack capacity) while generating solutions achieving the optimal reward (in relation to the original GMKP).

We show the proposed bi-GMKP algorithms can be adapted into GMKP heuristics to generate capacity feasible solutions, and also adapted into bi-GMKP heuristics to generate solutions with different combinations of rewards and maximum exceeded knapsack capacities\footnote{In this paper, the proposed methods that solve bi-GMKP with performance bounds are referred to as algorithms;
	proposed methods that solve GMKP (capacity-feasible) and bi-GMKP, both with no performance guarantees, are referred to as heuristics.}.
In an extensive computational study, algorithms and heuristics exhibit excellent performance overall.
In addition, we show that when capacities and weights are powers of the same positive integer, some of the proposed algorithms find optimal solutions.

This paper is organized as follows:
In \Cref{SEC:GMKP}, we define GMKP and present an \textit{Integer Programming} (IP) formulation.
In \Cref{SEC:bi-GMKP}, we define bi-GMKP, and propose three approximation algorithms in \Cref{SEC:approx-GMKP_LP,SEC:approx-GMKP_KP,SEC:approx-GMKP_mKP}.
\Cref{SEC:bi-GMKP_special} focuses on the algorithms' guarantees for some special cases of bi-GMKP\@.
In \Cref{SEC:heur}, we show how the proposed approximation algorithms for bi-GMKP can be used as heuristics for GMKP and bi-GMKP\@.
Finally, we test all algorithms and heuristics in an extensive computational study (\Cref{SEC:exp_design}), and present the conclusions in \Cref{SEC:conclusion}.

\section{GMKP Definition and IP Model}\label{SEC:GMKP}
In GMKP we are given a set of knapsacks \( {M=}{\{1,2,\dots,m\}}\), \(m\ge 2\), and a set of items \(N={\{1, 2, \dots, n\}}\).
Each knapsack \( {i\in M} \)  has a capacity \( c_i>0\), and each item \( {j\in N} \)  has a weight \(0<w_j\le\max_{i\in M}c_i\).
Items are partitioned intro groups \( {G_1\cup\dots\cup G_k=N}\), where \( K=\{1,2,\dots, k \} \), and each group \(G_l\), \(l\in K\), results in a reward \(p_l>0\) if assigned.
At least one group must have two or more items, if not, the instance would be an MKP\@.
\Cref{table:params} shows a summary of indices, sets, and parameters.
In a feasible assignment of items to knapsacks:
\begin{itemize}
	\item Each item is assigned to at most one knapsack.
	\item The capacity of a knapsack is not exceeded by the total weight of its assigned items.
	\item Whenever an item in a group is assigned to a knapsack, then all items in that group must be assigned to knapsacks.
\end{itemize}
The objective is to find a solution that maximizes the total reward from the groups assigned to knapsacks.
Without loss of generality we assume the following for the remainder of the paper:
\begin{itemize}
	\item \( \nexists l\in K\)  such that \( \sum_{j\in G_l}w_j>\sum_{i\in M}c_i\). Such groups cannot be feasibly assigned.
	\item  \( \min_{i\in M}c_i\ge\min_{j\in N} w_j\). If not, no item fits into the smallest knapsack and that knapsack can be removed.
\end{itemize}

\begin{table}[ht]\caption{Indices, sets, and parameters}
	\centering\label{table:params}
	\begin{tabular}{cl}
		\toprule
		\multicolumn{2}{l}{\textbf{Indices \& sets}} \\ \midrule
		\(i\in M=\{1,\dots,m\}\)
		 & index and set of knapsacks
		\\
		\(j\in N=\{1,\dots,n\}\)
		 & index and set of items
		\\
		\(l\in K=\{1,\dots,k\}\)
		 & index and set of groups
		\\ 
		\( {G_1\cup\dots\cup G_k=N}\)
		 & groups that partition the set of items
		\\     
		\midrule
		\multicolumn{2}{l}{\textbf{Parameters}}      \\ \midrule
		\(c_i\)
		 & capacity of knapsack \(i\in M\)
		\\
		\(w_j\)
		 & weight of item \(j\in N\)
		\\
		\(p_l\)
		 & reward of group \(l\in K\)
		\\    
		\( c_{\max} = \max\limits_{i\in M}c_i\)
		 & capacity of the largest knapsack
		\\
		\( w_{\max} = \max\limits_{j\in N}w_j\)
		 & weight of the heaviest item
		\\\bottomrule
	\end{tabular}
\end{table}

The following is an IP formulation for GMKP\@:
\begin{align*}
	x_{ij} & =\begin{cases}
		1, & \text{if item \( j\in N\)  is assigned to knapsack \( i\in M\)}
		\\
		0, & \text{otherwise}
	\end{cases}
	\\
	z_l    & =\begin{cases}
		1, & \text{if all items in group \( l\in K\)  are assigned to knapsacks}
		\\
		0, & \text{otherwise}
	\end{cases}
\end{align*}

\begin{definition}
	Given a GMKP instance:
	\begin{align}
		\textbf{IP-GMKP:\qquad}
		v^*=\max    & \sum_{l\in K} p_l z_l
		\label{eq:IP-GMKP-obj}                                                        \\
		\text{s.t.} & \sum_{j\in N} w_j x_{ij} \le c_i & i\in M\label{eq:IP-GMKP-cap} \\
		            & \sum_{i\in M} x_{ij}=z_l         & l\in K, j\in G_l
		\label{eq:IP-GMKP-eq}                                                         \\
		            & x_{ij}\in \{ 0, 1 \}             & i\in M,j\in N
		\nonumber                                                                     \\
		            & z_l\in \{ 0, 1 \}                & l \in K
		\nonumber
	\end{align}
\end{definition}

The objective function~\eqref{eq:IP-GMKP-obj} maximizes the total reward from the groups assigned to knapsacks.
Constraints~\eqref{eq:IP-GMKP-cap}  ensure that no knapsack capacity is exceeded.
Constraints~\eqref{eq:IP-GMKP-eq} guarantee that either all items within a group are assigned, or none are assigned at all; they also ensure that each item is assigned to at most one knapsack.

\section{bi-GMKP Definition and Approximation Algorithms}\label{SEC:bi-GMKP}

In bi-GMKP, we relax the capacity constraints~\eqref{eq:IP-GMKP-cap} and incorporate them as a second objective function
to minimize the maximum exceeded knapsack capacity, i.e.,
\begin{equation}
	\label{eq:bi-GMKP-obj}
	\min\max_{i\in M}  \left \{ \sum_{j\in N} w_j x_{ij} - c_i \right \}
\end{equation}
bi-GMKP is strongly \( \mathcal{NP}\)-hard, since GMKP is \citep{chen2018packing}.

\begin{definition}
	For \( 0<\alpha\le 1\)  and \( \beta \ge 0\),
	an algorithm  for bi-GMKP is an  \textbf{(\( \pmb\alpha, \pmb\beta \)) bi-criteria approximation algorithm} if any solution returned satisfies
	\begin{itemize}
		\item  \( \sum_{l\in K} p_l z_l \ge \alpha v^*\), where \( v^*\)  is the optimal objective value~\eqref{eq:IP-GMKP-obj} of the analogous GMKP\@.
		\item  \( \max_{i\in M}  \left \{ \sum_{j\in N} w_j x_{i j}- c_i\right \} \le \beta c_{\max} \).
	\end{itemize}
\end{definition}

In an optimal solution of GMKP, \(\alpha= 1\) and \(\beta=0\).
We propose three types of approximation algorithms for bi-GMKP, following a two step approach: group selection and item assignment.

\paragraph{Group selection:} Algorithms first select the groups to assign, focusing on maximizing rewards.
Each algorithm does this  by solving a  different relaxation of GMKP, which are:
\begin{enumerate}
	\item \Cref{alg:LP-GMKP}: LP-GMKP, a  \textit{Linear Programming} (LP) relaxation of IP-GMKP, where all binary variables \( x_{ij}\)  and \(z_l\) are replaced with continuous counterparts \( {0\le x_{ij}\le 1}\) and \( {0\le z_l\le 1}\)  respectively.
	      
	\item \Cref{alg:KP-GMKP}: KP-GMKP, a knapsack relaxation constructed by combining all capacity constraints~\eqref{eq:IP-GMKP-cap}.
	\item A \textit{multi-dimensional knapsack} (mKP) relaxation, which is a KP-GMKP with extra capacity constraints.
	      We define three versions of this algorithm:
	      \begin{itemize}
		      \item \Cref{alg:2mKP-GMKP}: A single extra capacity constraint (2mKP-GMKP).
		      \item \Cref{alg:3mKP_GMKP}: Two additional capacity constraints (3mKP-GMKP).
		      \item \Cref{alg:mKPD-GMKP}: Multiple extra capacity constraints based on a finite set \(D\subset\mathbb{R}_{>0}\) (\(\text{mKP}_\text{D}\)-GMKP); which generalizes 2mKP-GMKP and 3mKP-GMKP\@.
		            The theoretical results of the generalized version can be found in \Cref{APPENDIX:mKPD}.
	      \end{itemize}
\end{enumerate}
Note that the last two group selection methods sacrifice polynomial time, but are often faster to solve in computational experiments than the original GMKP\@.

\paragraph{Item assignment:}
Algorithms assign all items of the selected groups, focusing on minimizing the maximum exceeded knapsack capacity~\eqref{eq:bi-GMKP-obj}.
This is done greedily by sequentially assigning items (sorted from heavier to lightest) to the knapsacks with the most free capacity.

This item assignment sub-problem is equivalent to a parallel machine scheduling problem where each item corresponds to a job (durations are their weights), and each knapsack \( i\in M\)  corresponds to an identical parallel machine with earliest available time~\( c_{\max}-c_i\)
(i.e., the release time for machine \(i\)); aiming to minimize the total makespan.
This sub-problem is \( \mathcal{NP}\)-hard and has several polynomial time approximation algorithms \citep{lee1991parallel, lee2000note}.
\bigskip

\Cref{table:algorithms} summarizes the algorithms proposed and their approximation guarantees, for the general case of bi-GMKP and some special cases.
Note that all algorithms have \(\alpha= 1\), i.e., they achieve the optimal reward of the equivalent GMKP\@.
\begin{table}[ht]\caption{Summary of bi-GMKP algorithms' guarantees, \(\pmb{\alpha}\)=1 for all algorithms}\label{table:algorithms}
	\begin{tabularx}{\textwidth}{lccl}
		\toprule
		\textbf{Algorithm}
		 & \textbf{Relaxed Problem Solved}
		 & \(\pmb\beta\)
		\\\midrule
		\Cref{alg:LP-GMKP}
		 & LP-GMKP
		 & 2
		 & \multirow{5}{*}{General case}
		\\
		\Cref{alg:KP-GMKP}
		 & KP-GMKP
		 & 1
		\\
		\Cref{alg:2mKP-GMKP}
		 & 2mKP-GMKP
		 & \(1/2\)
		\\
		\Cref{alg:3mKP_GMKP}
		 & 3mKP-GMKP
		 & \(1/2\)
		\\
		\Cref{alg:mKPD-GMKP}
		 & \(\text{mKP}_\text{D}\)-GMKP, \(\left\{c_{\max}/2\right\}\subseteq D\)
		 & \(1/2\)
		\\\midrule
		\Cref{alg:LP-GMKP}
		 & LP-GMKP
		 & 2
		 & \multirow{5}{*}{Equal capacities}
		\\
		\Cref{alg:KP-GMKP}
		 & KP-GMKP
		 & 1
		\\
		\Cref{alg:2mKP-GMKP}
		 & 2mKP-GMKP
		 & \(1/2\)
		\\
		\Cref{alg:3mKP_GMKP}
		 & 3mKP-GMKP
		 & \(1/3\)
		\\
		\Cref{alg:mKPD-GMKP}
		 & \(\text{mKP}_\text{D}\)-GMKP, \(\left\{c_{\max}/2,c_{\max}/3\right\}\subseteq D\)
		 & \(1/3\)
		\\\midrule
		\multirow{2}{*}{\Cref{alg:2mKP-GMKP}}
		 & \multirow{2}{*}{2mKP-GMKP}
		 & \multirow{2}{*}{0}
		 & Equal capacities and
		\\
		 &
		 &
		 & items heavier than \(c_{\max}/2\)
		\\\midrule
		\multirow{3}{*}{\Cref{alg:KP-GMKP}}
		 & \multirow{3}{*}{KP-GMKP}
		 & \multirow{3}{*}{0}
		 & Equal capacities and
		\\
		 &
		 &
		 & capacities/weights are powers
		\\ &&& of the same positive integer
		\\\midrule
		\multirow{2}{*}{\Cref{alg:mKP'-GMKP}}
		 & \multirow{2}{*}{mKP'-GMKP}
		 & \multirow{2}{*}{0}
		 & Capacities/weights are powers
		\\ &&&of the same positive integer
		\\\bottomrule
	\end{tabularx}
\end{table}

\section{LP Based Approximation Algorithm for bi-GMKP}\label{SEC:approx-GMKP_LP}

\begin{definition}
	In a solution of an LP-GMKP instance, group \( {l\in K}\)  is a \textbf{partially assigned group} if \( {0<z_l<1}\).
\end{definition}

Some instances of LP-GMKP can have multiple optimal solutions, and there may be more than one partially assigned group in some of these solutions
(\Cref{prop:extreme_points} in \Cref{APPENDIX:LP-GMKP}).
To solve LP-GMKP, we propose a polynomial time greedy method which is guaranteed to return an optimal solution with at most one partially assigned group (\Cref{prop:LP-GMKP} and \Cref{cor:part_groups} in \Cref{APPENDIX:LP-GMKP});
this property is used in proving the approximation guarantee of \Cref{alg:LP-GMKP}.

\begin{algorithm}[ht]\caption{LP based approximation algorithm for bi-GMKP}\label{alg:LP-GMKP}\DontPrintSemicolon\SetAlgoNoLine
	\KwData{bi-GMKP instance.}
	\KwResult{\( x^a_{ij}\in \{ 0, 1 \} \), \( \forall i\in M,j\in N\);
	\( z^a_l\in \{ 0, 1 \} \), \( \forall l\in K\).}
	\nonl\textbf{Group selection:}\;
	Solve the corresponding LP-GMKP instance greedily (\Cref{alg:greedyLP} in \Cref{APPENDIX:LP-GMKP}), and get solution \( (x,z)\).\label{alg-line:LP-GMKP_LP}\;
	\( z^a_l\gets 1\), \( \forall l\in K\) such that \(z_l>0\).\label{alg-line:LP-GMKP_LP2}
	\;
	\nonl\textbf{Item assignment:}\;
	\( x^a_{ij}\gets 0\), \( \forall i\in M,j\in  N\).
	\;
	Let \( N^\text{s}=\bigcup\limits_{l:z^a_l=1} G_l\),  be the set containing all items of the selected groups.
	\;
	\ForEach{\( j^\text{s}\in N^\text{s}\) }{
	\( x^a_{ij^\text{s}}\gets 1\), where \( i=\argmin\limits_{i\in M} \left\{\sum\limits_{j\in N} w_j x_{i j}- c_i  \right\} \).\label{alg-line:LP-GMKP_items}
	}
	Return solution \( (x^a,z^a)\).
	\;
\end{algorithm}

\begin{theorem}\label{theo:approx_LP-GMKP}
	\Cref{alg:LP-GMKP} {\normalfont (i)} is a (1,2)-approximation algorithm, and {\normalfont (ii)} runs in polynomial time.
		{\normalfont (iii)} This is a tight approximation.
\end{theorem}

\proof{Proof of \Cref{theo:approx_LP-GMKP}}:
(i)
Let \( (x^a,z^a)\)  be the solution obtained by the algorithm.
Guarantee \(\alpha\ge1\) is trivial, since LP-GMKP is a linear relaxation of GMKP\@.

Now we prove guarantee \( \beta\le 2\).
The algorithm solves an LP-GMKP instance and selects all groups that are entirely or partially assigned.
Then it greedily assigns the items of the selected group, one by one, to the knapsack with the least exceeded (or most remaining) capacity.
Suppose by contradiction that as items are assigned to knapsacks, the capacity is exceeded by more than \( 2c_{\max}\) after assigning some item \(j\).
This means that every knapsack has its capacity exceeded by more than \( c_{\max}\) (if not, item \(j\) would be assigned to the knapsack with the least exceeded capacity, without exceeding any knapsacks capacity by more than \( 2c_{\max}\)), therefore,
\begin{equation}\label{eq:proof_LP-GMKP}
	\sum_{i\in M}\sum_{j\in N}w_j x^a_{i j}>\sum_{i\in M}(c_i+c_{\max})\ge 2\sum_{i\in M}c_i
\end{equation}
On the other hand, the partially assigned group (at most one from \Cref{cor:part_groups} in \Cref{APPENDIX:LP-GMKP}) cannot weigh more than the sum of all capacities, and the same goes for all other selected groups  together.
Therefore \( {\sum_{i\in M}\sum_{j\in N}w_j x^a_{i j}\le 2\sum_{i\in M}c_i}\),  contradicting~\eqref{eq:proof_LP-GMKP}.

(ii) \Cref{alg:LP-GMKP} runs in polynomial time
\( {\mathcal{O}(k\log(k) +n+ m\log(m))}\); where
\( k\log(k)\) + \(n\) corresponds to the run time of greedily solving LP-GMKP (\Cref{prop:LP-GMKP}  in \Cref{APPENDIX:LP-GMKP}),
and
\( { m\log(m)}\) corresponds to the run time of the greedy assignment of items  (\cref{alg-line:LP-GMKP_items}), by sorting and keeping the list updated.

(iii)
See \hyperref[eg:LP]{Examples for  \Cref{theo:approx_LP-GMKP}} in \Cref{APPENDIX:tight}.
\Halmos
\endproof

\section{KP Based Pseudo-Polynomial Time Approximation Algorithm for bi-GMKP}\label{SEC:approx-GMKP_KP}
\begin{definition}\label{def:KP-GMKP}
	Given a GMKP instance:
	\begin{align}
		\textbf{KP-GMKP:\qquad}\max & \sum_{l\in K} p_l z_l
		\tag{\ref*{eq:IP-GMKP-obj}}                                                                                              \\
		\text{s.t.}                 & \sum_{l\in K}\sum_{j\in G_l} w_j z_l \le \sum_{i\in M} c_i\label{eq:KP-GMKP-cap}           \\
		                            & z_l\in \{ 0, 1 \}                                                                & l \in K
		\nonumber
	\end{align}
\end{definition}

\begin{lemma}\label{lemma:KP-GMKP_relax}
	KP-GMKP is a relaxation of the corresponding GMKP\@.
\end{lemma}
\proof{Proof of \Cref{lemma:KP-GMKP_relax}}:
Consider IP-GMKP\@.
Adding all capacity constraints~\eqref{eq:IP-GMKP-cap} and replacing \( \sum_{i\in M} x_{ij}\) with \( z_l\) for \(j\in G_l\), we get
\[
	\sum_{i\in M} c_i
	\ge \sum_{i\in M}\sum_{j\in N} w_j x_{ij}
	= \sum_{j\in N}w_j \left[\sum_{i\in M} x_{ij} \right]
	= \sum_{l\in K}\sum_{j\in G_l}w_j \left[\sum_{i\in M} x_{ij} \right]
	= \sum_{l\in K}\sum_{j\in G_l}w_j z_l
\]
This is constraint~\eqref{eq:KP-GMKP-cap}.
\Halmos
\endproof

\begin{algorithm}[ht]\DontPrintSemicolon
	\caption{KP based approximation algorithm for bi-GMKP}\label{alg:KP-GMKP}
	\KwData{bi-GMKP instance.}
	\KwResult{\( x^a_{ij}\in \{ 0, 1 \} \), \( \forall i\in M,j\in N\);
	\( z^a_l\in \{ 0, 1 \} \), \( \forall l\in K\).}
	\nonl Run \Cref{alg:LP-GMKP}, changing \cref{alg-line:LP-GMKP_LP,alg-line:LP-GMKP_LP2} with the following:
	\\\nonl Solve the corresponding KP-GMKP instance, and get solution \( z^a\).\label{alg-line:KP-KP}\;
\end{algorithm}

\begin{theorem}\label{theo:approx_KP-GMKP}
	\Cref{alg:KP-GMKP} {\normalfont (i)} is a (1,1)-approximation algorithm, and
		{\normalfont (ii)} runs in pseudo-polynomial time.
		{\normalfont (iii)} This is a tight approximation.
\end{theorem}

\proof{Proof of \Cref{theo:approx_KP-GMKP}}:
(i)
Guarantee \( \alpha\ge1\) is trivial since KP-GMKP is a relaxation of GMKP (\Cref{lemma:KP-GMKP_relax}).
Now we prove guarantee \( \beta\le 1\).
The groups selected by KP-GMKP do not exceed the total knapsack capacity. Therefore, during the greedy item assignment stage, items are always assigned to a knapsack with free capacity (the capacity might be exceeded once the item gets assigned).
Thus, the capacity of a knapsack cannot be exceeded by more than \( w_{\max}\le c_{\max}\).

(ii) The algorithm runs in pseudo-polynomial time
\[{\mathcal{O}\left(m\log(m) + n\sum_{i\in M} c_i\right)}\]
where \(m\log(m)\) corresponds to the greedy assignment of items, and \(n\sum_{i\in M} c_i\) to the pseudo-polynomial solution time of KP \citep{horowitz1974computing}.

(iii)
See \hyperref[eg:KP]{Examples for  \Cref{theo:approx_KP-GMKP}} in \Cref{APPENDIX:tight}.
\Halmos
\endproof

\section{2mKP \& 3mKP Based Pseudo-Polynomial Time Approximation Algorithm for bi-GMKP}\label{SEC:approx-GMKP_mKP}

For \( d>0\)  define the following function \( f_d:\mathbb{R}_{> 0}\to\mathbb{Z}_{\ge 0}\)
\begin{equation*}
	f_d(y)= \max
	\left\{q\in \mathbb{Z} : q< \frac{y}{d} \right\}
	=
	\begin{cases}
		\frac{y}{d}-1,                        & \text{if }\frac{y}{d}\in\mathbb{Z},
		\\
		\left\lfloor\frac{y}{d}\right\rfloor, & \text{if }\frac{y}{d}\notin\mathbb{Z}
	\end{cases}
\end{equation*}
Intuitively, function \( f_d(y)\)  represents the number of times items slightly larger than  \( d \)  completely fit into \( y\).
\( \lfloor \cdot \rfloor \)  denotes the integer part or floor function.

\begin{definition}
	Given a GMKP instance:
	\begin{align}
		\textbf{2mKP-GMKP:\qquad}\max & \sum_{l\in K} p_l z_l
		\tag{\ref*{eq:IP-GMKP-obj}}                                                                                                                         \\
		\text{s.t.}                   & \sum_{l\in K}\sum_{j\in G_l} w_j z_l\le \sum_{i\in M} c_i
		\tag{\ref*{eq:KP-GMKP-cap}}
		\label{eq:KP-GMKP-cap2}                                                                                                                             \\
		                              & \sum_{l\in K}\sum_{j\in G_l} f_{\frac{c_{\max}}{2}}(w_j) z_l\le \sum_{i\in M} f_{\frac{c_{\max}}{2}}(c_i) & 
		\label{eq:2mKP-GMKP-constr}                                                                                                                         \\
		                              & z_l\in \{ 0, 1 \}                                                                                         & l \in K
		\nonumber
	\end{align}
\end{definition}

Constraint~\eqref{eq:2mKP-GMKP-constr} is a valid cut for GMKP, i.e., avoids some infeasible GMKP solutions that are feasible in KP-GMKP\@. For example, consider two knapsacks with equal capacities 1, and only one group with three items that weigh 0.6 each. The group cannot be assigned without exceeding the capacity of some knapsack, but KP-GMKP would still select the group.
Adding constraint~\eqref{eq:2mKP-GMKP-constr} namely,
\(
3f_{\frac{1}{2}}(0.6)z_1=3z_1\le 2f_{\frac{1}{2}}(1)=2,
\)
avoids selecting the group.

\begin{lemma}\label{lemma:2mKP-GMKP_relax}
	A 2mKP-GMKP instance is a relaxation of its corresponding GMKP instance.
\end{lemma}
\proof{Proof of \Cref{lemma:2mKP-GMKP_relax}}:
KP-GMKP  is a relaxation of GMKP (\Cref{lemma:KP-GMKP_relax}),
therefore it suffices to show that constraint~\eqref{eq:2mKP-GMKP-constr} is satisfied by all feasible solutions of GMKP\@.
For \({d=c_{\max}/2}\)
\begin{align}\label{eq:fd_claim}
	\sum_{j\in N} f_d(w_j)x_{ij}\le f_d(c_i)
	 &  & i\in M
\end{align}
holds for all GMKP solutions, because capacity constraints~\eqref{eq:IP-GMKP-cap} are satisfied by GMKP solutions and \(f_d\) is supper-additive (i.e., \( f_d(y_1)+f_d(y_2)\le f(y_1+y_2)\), for all \( y_1,y_2\ge 0\)).
By adding all constraints~\eqref{eq:fd_claim} together we get
\begin{align*}
	\sum_{i\in M}f_d(c_i)
	\ge
	\sum_{i\in M}\sum_{j\in N} f_d(w_j)x_{ij}
	= \sum_{l\in K}\sum_{j\in G_l}f_d(w_j) \left[\sum_{i\in M} x_{ij} \right]
	= \sum_{l\in K}\sum_{j\in G_l}f_d(w_j) z_l
\end{align*}
This is constraint~\eqref{eq:2mKP-GMKP-constr} when we substitute \({d=c_{\max}/2}\).

\Halmos
\endproof

\begin{algorithm}[ht]\DontPrintSemicolon
	\caption{2mKP based approximation algorithm for bi-GMKP}\label{alg:2mKP-GMKP}
	\KwData{bi-GMKP instance.}
	\KwResult{\( x^a_{ij}\in \{ 0, 1 \} \), \( \forall i\in M,j\in N\);
	\( z^a_l\in \{ 0, 1 \} \), \( \forall l\in K\).}
	\nonl Run \Cref{alg:LP-GMKP}, changing \cref{alg-line:LP-GMKP_LP,alg-line:LP-GMKP_LP2} with the following:
	\\\nonl Solve the corresponding 2mKP-GMKP instance, and get solution \( z^a\).\;
\end{algorithm}

\begin{theorem}\label{theo:approx_mKP-GMKP}
	\Cref{alg:2mKP-GMKP}
	{\normalfont(i)}
	is a \( \left(1, 1/2\right)\)-approximation algorithm, and
		{\normalfont(ii)} runs in pseudo-polynomial time.
		{\normalfont(iii)} This is a tight approximation.
\end{theorem}

\proof{Proof of \Cref{theo:approx_mKP-GMKP}}:
(i)
Guarantee \(\alpha\ge1\) is trivial, since 2mKP-GMKP is a relaxation of GMKP (\Cref{lemma:2mKP-GMKP_relax}).
We now prove guarantee \( \beta\le 1/2\).
The groups selected by 2mKP-GMKP do not exceed the total knapsack capacity. Therefore, during the greedy item assignment stage, items are always assigned to a knapsack with free capacity
and thus
knapsack capacities can  be exceeded by more than \( c_{\max}/2\) only when items heavier than \( c_{\max}/2\) are assigned.
Constraint~\eqref{eq:2mKP-GMKP-constr} implies that when an item \(j\) with \(w_j\ge c_{\max}/2\) is assigned, there must exist a knapsack with capacity larger than \( c_{\max}/2\), and item \(j\) is the first item assigned to this knapsack.
Hence, no knapsacks' capacity can be exceeded by more than \( c_{\max}/2\).

(ii) The algorithm runs in pseudo-polynomial time
\[ {\mathcal{O}\left(m\log(m) + n\left(\sum_{i\in M} c_i \right)\left(\sum_{i\in M} f_{\frac{c_{\max}}{2}}(c_i) \right)\right)}\]
where \(m\log(m)\) corresponds to the greedy assignment of items, and the second term corresponds to the pseudo-polynomial solution time of mKP \citep{freville2004multidimensional}.

(iii)
See \hyperref[eg:mKP]{Examples for  \Cref{theo:approx_mKP-GMKP}} in \Cref{APPENDIX:tight}.
\Halmos
\endproof

\begin{definition}
	Given a GMKP instance:
	\begin{align}
		\textbf{3mKP-GMKP:\qquad}\max & \sum_{l\in K} p_l z_l
		\tag{\ref*{eq:IP-GMKP-obj}}                                                                                                                         \\
		\text{s.t.}                   & \sum_{l\in K}\sum_{j\in G_l} w_j z_l\le \sum_{i\in M} c_i
		\tag{\ref*{eq:KP-GMKP-cap}} \label{eq:KP-GMKP-cap2.2}                                                                                               \\
		                              & \sum_{l\in K}\sum_{j\in G_l} f_{\frac{c_{\max}}{2}}(w_j) z_l\le \sum_{i\in M} f_{\frac{c_{\max}}{2}}(c_i) & 
		\tag{\ref*{eq:2mKP-GMKP-constr}}\label{eq:2mKP-GMKP-constr2}                                                                                        \\
		                              & \sum_{l\in K}\sum_{j\in G_l} f_{\frac{c_{\max}}{3}}(w_j) z_l\le \sum_{i\in M} f_{\frac{c_{\max}}{3}}(c_i) & 
		\label{eq:3mKP-GMKP-constr}                                                                                                                         \\
		                              & z_l\in \{ 0, 1 \}                                                                                         & l \in K
		\nonumber
	\end{align}
\end{definition}

\begin{algorithm}[ht]\DontPrintSemicolon
	\caption{3mKP based approximation algorithm for bi-GMKP}\label{alg:3mKP_GMKP}
	\KwData{bi-GMKP instance.}
	\KwResult{\( x^a_{ij}\in \{ 0, 1 \} \), \( \forall i\in M,j\in N\);
	\( z^a_l\in \{ 0, 1 \} \), \( \forall l\in K\).}
	\nonl Run \Cref{alg:LP-GMKP}, changing \cref{alg-line:LP-GMKP_LP,alg-line:LP-GMKP_LP2} with the following:
	\\\nonl Solve the corresponding 3mKP-GMKP instance, and get solution \( z^a\).\;
\end{algorithm}
\Cref{alg:3mKP_GMKP} has the same guarantees as \Cref{alg:2mKP-GMKP} for the general case of bi-GMKP\@; they are both \( \left(1, 1/2\right)\)-approximation algorithms.
Although, \Cref{alg:3mKP_GMKP} has a better guarantee for some special case as seen in \Cref{sSEC:approx-equal_cap}.

A generalization of 2mKP-GMKP and 3mKP-GMKP can be found in \Cref{APPENDIX:mKPD}.
There, \Cref{theo:approx_mKP-GMKP2} shows that additional constraints in the form of constraints~\eqref{eq:2mKP-GMKP-constr} (with different \(d>0\), \(d\neq c_{\max}/2\) values) do not improve the \(\beta\le 1/2\) guarantee of \Cref{theo:approx_mKP-GMKP}, even when all possible constraints of such form are included (\Cref{cor:bi-GMKP2} in \Cref{APPENDIX:mKPD}).

\section{Special Cases of bi-GMKP}\label{SEC:bi-GMKP_special}

In this section we study two special cases of bi-GMKP, namely, equal capacity knapsacks (\Cref{sSEC:approx-equal_cap}), and when item weights and knapsack capacities are powers of the same positive integer (\Cref{sSEC:approx-Po2}).

\subsection{Equal Capacity Knapsacks}\label{sSEC:approx-equal_cap}

\begin{corollary}\label{cor:approx_2KP-GMKP_w}
	When all knapsacks have equal capacities and all items are heavier than \(c_{\max}/2\),
	then
	\Cref{alg:2mKP-GMKP} returns an optimal solution of its corresponding GMKP instance.
\end{corollary}

\proof{Proof of \Cref{cor:approx_2KP-GMKP_w}}:
To satisfy constraint~\eqref{eq:2mKP-GMKP-constr}, there exists a knapsack for each item assigned.
Hence, no capacity is exceeded.
\Halmos
\endproof

\begin{theorem}\label{theo:approx_3mKP}
	When all knapsacks have equal capacities,
	\Cref{alg:3mKP_GMKP}
	{\normalfont (i)} is a \( \left(1,1/3\right)\)-approximation algorithm, and
		{\normalfont (ii)}  runs in pseudo-polynomial time.
		{\normalfont (iii)} This is a tight approximation.
\end{theorem}

\proof{Proof of \Cref{theo:approx_3mKP}	}:
(i)
Guarantee \(\alpha\ge1\) is trivial, since 3mKP-GMKP is a relaxation of GMKP (analogous to \Cref{lemma:2mKP-GMKP_relax}).
We now prove guarantee \( \beta\le 1/3\).
The groups selected by 3mKP-GMKP do not exceed the total knapsack capacity. Therefore, during the greedy item assignment stage, items are always assigned to a knapsack with free capacity,
so any item that weighs \(c_{\max}/3\) or less cannot exceed the capacity by more than \(c_{\max}/3\) when assigned.
Constraint~\eqref{eq:2mKP-GMKP-constr2} ensures that the number of selected items larger than \(c_{\max}/2\) cannot exceed the number of knapsacks. Hence, the algorithm always assigns an item heavier than \(c_{\max}/2\) to an empty knapsack.

The interesting case is assigning an item \( j\in N\) with \( c_{\max}/3<w_j\le c_{\max}/2\) to a knapsack \( i\in M\).
Suppose the algorithm assigns such an item \(j\) to a knapsack \(i\) and exceeds its capacity; we will show that the capacity cannot be exceeded by more than \(c_{\max}/3\).

\begin{itemize}
	\item If knapsack \( i\) is empty, its capacity cannot be exceeded.
	      
	\item If knapsack \( i\)  has one item assigned previously, such item must weigh more than \(5c_{\max}/6\) for the knapsack capacity to be exceeded by more than \( c_{\max}/3\) after the assignment of item \(j\).
	      This means that, after assigning item \(j\), this knapsack's contribution to the \textit{left-hand side} (lhs) of constraint~\eqref{eq:3mKP-GMKP-constr} would be 3; while the contribution to the \textit{right-hand side} (rhs) of each knapsack would be 2.
	      Therefore, to satisfy the constraint, there must exist another knapsack whose contribution to the lhs is either 1 or 0.
	      0 is not possible since the knapsack must have at least one item at least as heavy as item \(j\).
	      If the contribution to the lhs is 1, it means the knapsack has only one item which is lighter than \(2c_{\max}/3\), but in such a case, the capacity cannot be exceeded by more than \(c_{\max}/3\).
	      
	\item If knapsack \( i\)  has two or more items, then  this knapsack's contribution to the lhs of constraint~\eqref{eq:3mKP-GMKP-constr} would be at least 3 (after assigning item \(j\)).
	      From the previous part, this contradicts the fact that constraint~\eqref{eq:3mKP-GMKP-constr} is satisfied.
\end{itemize}

(ii) The algorithm runs in pseudo-polynomial time
\[\mathcal{O}\left(
	m\log(m) + n\left(\sum_{i\in M} c_i \right)
	\left(\sum_{i\in M} f_{\frac{c_{\max}}{2}}(c_i) \right)
	\left(\sum_{i\in M} f_{\frac{c_{\max}}{3}}(c_i) \right)
	\right)\]
where \(m\log(m)\) corresponds to the greedy assignment of items, and the second term corresponds to the pseudo-polynomial solution time of mKP \citep{freville2004multidimensional}.

(iii)
See \hyperref[eg:3mKP]{Examples for  \Cref{theo:approx_3mKP}} in \Cref{APPENDIX:tight}.
\Halmos
\endproof
\bigskip

\Cref{theo:approx_mKP-GMKP3} in \Cref{APPENDIX:mKPD}  shows that if a finite number of constraints in the form of constraint~\eqref{eq:3mKP-GMKP-constr} (with different \(d>0\), \(d\neq c_{\max}/2\), \(d\neq c_{\max}/3\) values) are added to 3mKP-GMKP, this does not improve the \(\beta\le 1/3\) guarantee of \Cref{theo:approx_3mKP}.

\begin{corollary}\label{cor:approx_tight}
	When all knapsacks have equal capacities, all approximation guarantees in \Cref{theo:approx_LP-GMKP,theo:approx_KP-GMKP,theo:approx_mKP-GMKP,theo:approx_3mKP} are tight.
\end{corollary}
\proof{Proof of \Cref{cor:approx_tight}}:
All tight examples used in the proofs have equal capacities.
\Halmos
\endproof

\subsection{Capacities and Weights are Powers of Integer
	\texorpdfstring{\(\boldsymbol a\)}{a}
}\label{sSEC:approx-Po2}
\begin{corollary}\label{cor:approx_KP-GMKP_Po2}
	When all capacities and weights are powers of integer \(a>0\),
	and knapsacks have equal capacities,
	\Cref{alg:KP-GMKP} returns an optimal solution of its corresponding GMKP instance.
\end{corollary}

\proof{Proof of \Cref{cor:approx_KP-GMKP_Po2}}:
Suppose by contradiction that the algorithm assigns some item \(j\in N\) to knapsack \(i\in M\), exceeding its capacity.
The weight and capacity can be expressed, respectively, as \(w_j=a^q\) and \(c_i=a^r\), for some \({q,r\in\mathbb{Z}_{> 0}}, r\ge q\).
The free capacity in knapsack \(i\), before assigning item \(j\),  can be expressed as \(a^r-sa^q\) for some \({s\in\mathbb{Z}_{\ge 0}}\) (recall heavier items are assigned first).
To exceed the knapsack's capacity, \(0<a^r-sa^q<a^q\) must hold,
which implies \({s<a^{r-q}<s+1}\); not possible for \(r\ge q\) since \(s\) is integer.
\Halmos
\endproof

\begin{definition}
	Given a GMKP instance:
	\begin{align}
		\textbf{mKP'-GMKP:\qquad}\max & \sum_{l\in K} p_l z_l
		\tag{\ref*{eq:IP-GMKP-obj}}                                                                         \\
		\text{s.t.}                   & \sum_{l\in K}\sum_{j\in G_l} w_j z_l\le \sum_{i\in M} c_i
		\tag{\ref*{eq:KP-GMKP-cap}} \label{eq:IP-GMKP-obj.2}                                                \\
		                              & 
		\sum_{l\in K}\sum_{j\in G_l} \left\lfloor \frac{w_j}{d}\right\rfloor z_l\le \sum_{i\in M} \left\lfloor \frac{c_i}{d}\right\rfloor
		                              & d\in D
		= \left\{w_j : j\in N, w_j>\min_{i\in M}c_i  \right\}
		\label{eq:mKP'-GMKP-constr}                                                                         \\
		                              & z_l\in \{ 0, 1 \}                                         & l \in K
		\nonumber
	\end{align}
\end{definition}

\begin{algorithm}[ht]\DontPrintSemicolon
	\caption{Alternative mKP based approximation algorithm for bi-GMKP}\label{alg:mKP'-GMKP}
	\KwData{bi-GMKP instance.  }
	\KwResult{\( x^a_{ij}\in \{ 0, 1 \} \), \( \forall i\in M,j\in N\);
	\( z^a_l\in \{ 0, 1 \} \), \( \forall l\in K\).}
	\nonl Run \Cref{alg:LP-GMKP}, changing \cref{alg-line:LP-GMKP_LP,alg-line:LP-GMKP_LP2} with the following:
	\\\nonl Solve the corresponding mKP'-GMKP instance, and get solution \( z^a\).\;
\end{algorithm}

\begin{theorem}\label{theo:approx_mKP'-GMKP}
	When all capacities and weights are powers of integer \(a>0\),
	\Cref{alg:mKP'-GMKP} returns an optimal solution of its corresponding GMKP instance.
\end{theorem}

\proof{Proof of \Cref{theo:approx_mKP'-GMKP}}:
Since mKP'-GMKP is a relaxation of its respective GMKP instance (analogous to \Cref{lemma:2mKP-GMKP_relax}), it suffices to show that the solution returned by the algorithm does not exceed any knapsack capacity.

Suppose by contradiction that the algorithm assigns item \(j\in N\) to some knapsack, exceeding its capacity.
Note that all knapsacks \(i\in M\) with \(c_{i}\ge w_j\) are completely full;
if not, item \(j\) would fit entirely in some knapsack's free capacity, since capacities and weights are powers of \(a\) (recall heavier items are assigned first).
This also implies that \(w_j>\min_{i\in M}c_i\) must hold for item \(j\) to exceed the capacity of a knapsack (if not, all knapsacks are full which contradicts constraint~\eqref{eq:IP-GMKP-obj.2}).

When \(w_j>\min_{i\in M}c_i\),
consider constraint~\eqref{eq:mKP'-GMKP-constr} for \( d=w_j\in D\).
Each knapsack \(i\in M\) such that \(c_{i}\ge w_j\) is full and contributes to the lhs and rhs equally.
Other knapsacks with capacities smaller than \(w_j\) contribute 0 to the rhs, but item \(j\) contributes 1 to the lhs.
Thus, constraint~\eqref{eq:mKP'-GMKP-constr} for \( d=w_j\in D\) would not be satisfied, a contradiction.
\Halmos
\endproof

\section{Heuristics for GMKP and bi-GMKP}\label{SEC:heur}

\Cref{alg:LP-GMKP}--\ref{alg:3mKP_GMKP} and~\ref{alg:mKPD-GMKP} can exceed knapsack capacities, and their solutions can be improved through local search heuristics, e.g.,
jump and swap operations:
a jump operation consists of moving one item to another knapsack, and a swap operation consists of exchanging the assignments of two different items.
A swap-optimal solution is such that there are no swap or jump operations that improve the solution.

This approach is widely used to solve parallel-machine scheduling problems \citep{schuurman2007performance}, and swap-optimal solutions can be found in polynomial time \citep{finn1979linear}.
Therefore, given a bi-GMKP solution, it is possible to obtain a swap-optimal solution in polynomial time.

\hyperref[heur:binary-GMKP2]{Heuristic~\ref*{heur:binary-GMKP}} is a binary-search GMKP heuristic that is trivially guaranteed to stop, and obtains a capacity-feasible solution.
In the worst-case scenario, the obtained solution has no groups nor items assigned.
The heuristic's logic is to find a feasible solution utilizing the capacity as much as possible, by exploring the solution space through binary search.
Note \hyperref[heur:binary-GMKP2]{Heuristic~\ref*{heur:binary-GMKP}} incorporates a swap-optimal improvement, and assumes without loss of generality that all capacities and weights are integer.

\begin{heuristic}[ht]\DontPrintSemicolon
	\caption{Binary-search GMKP heuristic}\label[heuristic]{heur:binary-GMKP}
	\KwData{GMKP instance with integer capacities and weights.}
	\KwResult{\( x^h_{ij}\in \{ 0, 1 \} \), \( \forall i\in M,j\in N\);
	\( z^h_l\in \{ 0, 1 \} \), \( \forall l\in K\).}
	\( left\gets 0\) \;\label{heur:binary-GMKP2}
	\( right\gets \sum_{i\in M}c_i\) \;
	\( (x^h,z^h)\gets (\textbf{0},\textbf{0})\), i.e., solution with no items nor groups assigned.\;
	\While{\( l\le r\) }{
		\( TotalCapacity\gets \left\lfloor\frac{left+right}{2}\right \rfloor \) \;
		Solve the corresponding bi-GMKP instance with a slightly modified \Cref{alg:LP-GMKP}--\ref{alg:3mKP_GMKP} or~\ref{alg:mKPD-GMKP}, and get solution
		\( (x^a,z^a)\).
		The modified version of each algorithm consists in replacing the rhs of the capacity constraints, \( \sum_{i\in M}c_i\), with the current value of \( TotalCapacity\).\label{alg-line:binary-GMKP}
		\;
		Do a swap-optimal improvement on solution \( (x^a,z^a)\).
		\;
		\uIf{maximum exceeded knapsack capacity of solution
			\( (x^a,z^a)\)  is 0 or less}{
			\( left\gets TotalCapacity+1\) \;
			\( (x^h,z^h)\gets (x^a,z^a)\) \;
		}
		\Else{
			\( right\gets TotalCapacity-1\)
		}
	}
	Return solution \( (x^h,z^h)\).\;
\end{heuristic}

A modified versions of \Cref{alg:LP-GMKP}--\ref{alg:3mKP_GMKP} and~\ref{alg:mKPD-GMKP}, as seen in \hyperref[heur:bi-GMKP2]{Heuristic~\ref*{heur:bi-GMKP}}, can run with different \textit{TotalCapacity} values to obtain bi-GMKP solutions.
If we run \hyperref[heur:bi-GMKP2]{Heuristic~\ref*{heur:bi-GMKP}} for several alternative \textit{TotalCapacity} values, we no longer maintain the \(\alpha,\beta\) guarantees but can identify a set of non-dominated solutions (in terms of bi-criteria: rewards vs.~maximum exceeded knapsack capacity).

\begin{heuristic}[ht]\DontPrintSemicolon
	\caption{bi-GMKP heuristic}\label[heuristic]{heur:bi-GMKP}
	\KwData{bi-GMKP instance, \( TotalCapacity>0\).  }
	\KwResult{\( x^h_{ij}\in \{ 0, 1 \} \), \( \forall i\in M,j\in N\);
	\( z^h_l\in \{ 0, 1 \} \), \( \forall l\in K\).}
	Solve the corresponding bi-GMKP instance with a slightly modified \Cref{alg:LP-GMKP}--\ref{alg:3mKP_GMKP} or~\ref{alg:mKPD-GMKP}, and get solution
	\( (x^h,z^h)\).
	The modified version of each algorithm consists in replacing the rhs of the capacity constraints, \( \sum_{i\in M}c_i\), with the value of \( TotalCapacity\).\label{heur:bi-GMKP2}
	\;
	Do a swap-optimal improvement on solution \( (x^h,z^h)\).
	\;
	Return solution \( (x^h,z^h)\).\;
\end{heuristic}

\section{Computational Study}\label{SEC:exp_design}

We created an extensive set of randomly generated GMKP instances to test the performance of bi-GMKP approximation algorithms (\Cref{alg:LP-GMKP}--\ref{alg:3mKP_GMKP} and~\ref{alg:mKPD-GMKP}),
GMKP heuristics  (\hyperref[heur:binary-GMKP2]{Heuristic~\ref*{heur:binary-GMKP}}), and bi-GMKP heuristics (\hyperref[heur:bi-GMKP2]{Heuristic~\ref*{heur:bi-GMKP}}).

The implementation was done on Python 3.6.5, and solver Gurobi 8.0.0 (with default settings) to solve IP models: IP-GMKP, KP-GMKP, 2mKP-GMKP, 3mKP-GMKP, and \(\text{mKP}_\text{D}\)-GMKP\@.
\ifUseOnlineSupplements
	Instances used in the experimental evaluation can be found in the online supplements.
\fi
All bi-GMKP algorithms tested are abbreviated as follows
\begin{itemize}
	\item LP:\ \Cref{alg:LP-GMKP} (sub-problem LP-GMKP).
	\item KP:\ \Cref{alg:KP-GMKP} (sub-problem KP-GMKP).
	\item 2mKP:\: \Cref{alg:2mKP-GMKP} (sub-problem 2mKP-GMKP).
	\item 3mKP:\: \Cref{alg:3mKP_GMKP} (sub-problem 3mKP-GMKP).
	\item 100mKP:\: \Cref{alg:mKPD-GMKP}
	      for \(D=\left\{100/2,100/3,\dots,100/100\right\}\)
	      (sub-problem \(\text{mKP}_\text{D}\)-GMKP).
	\item Best: Selects the best solution between all previous algorithms.
	\item IP:\ Solves IP-GMKP with Gurobi, limiting each running time to three hours.
\end{itemize}
Experiments ran on a computer cluster with over 500 nodes.
When solving a specific instance, all algorithms solved the instance in the same node to ensure a fair comparison of running time between algorithms.

\subsection{Instance Generation}\label{sSEC:instances}

All instances generated have integer weights and all knapsacks have an equal capacity of 100.
We set the reward of each group equal to the total weight of the items in that group; some alternative reward values were also tested (see \Cref{APPENDIX:results}).
We focused the computational study on instances with equal knapsack capacities, since~\Cref{alg:LP-GMKP}--\ref{alg:3mKP_GMKP} have different guarantees on the maximum exceeded knapsack capacity (\(\beta\le 2, 1, 1/2,1/3\) respectively).

We generated 3,000 instances from a 6 dimensional maximum projection Latin hypercube design \citep{joseph2015maximum} by using the \textit{R} package \textit{MaxPro}; based on a simulated annealing algorithm  \citep{maxpro}.
The parameters of each instance are based on six random variables uniformly distributed between 0 and 1, where each random variable is later transformed to:
\begin{enumerate}[(i)]
	\item \( m\): Uniformly sets the number of knapsacks from 2 to 100.
	\item  \( w_\text{split}\): The weight difference between the heaviest and lightest items.	Uniformly choose a random integer from 1 to 99.
	\item  \( w_{\min}\):  The weight of the lightest item.
	      Uniformly choose a random integer from 1 to \( \min(50\), \(100-w_\text{split})\). \( w_{\min}\) is capped at \(c_{\max}/2=50\), to avoid instances solved to optimality by \Cref{alg:2mKP-GMKP} (\Cref{cor:approx_2KP-GMKP_w}).
	\item  \( w_\text{mode}\):  The desired mode of the item weights.
	      Uniformly choose a random integer from \( w_{\min}\)  to \( w_{\max}=w_{\min}+w_\text{split}\).
	\item   \( r_\text{load}\):  The desired load-ratio \( (\sum_{j\in N}w_j)/(\sum_{i\in M}c_i)\) for the knapsacks.
	      Uniformly choose a real number from \( 1\)  to \( 20\).
	      
	\item  \(r_\text{conc}\):  The desired concentration-ratio \(  1-k/n\) (correlated with the average number of items in a group, \(n/k\)).
	      Uniformly choose a real number from \( 0\)  to \( 1\).
\end{enumerate}
Once these parameters are defined for an instance, items and groups are created as follows:
\begin{itemize}
	\item Generate two items of weights \( w_{\min}\)  and \( w_{\max}\).
	\item Generate items until \( (\sum_{j\in N}w_j)/(\sum_{i\in M}c_i)\)  exceeds the desired load-ratio.
	      Each weight is randomly chosen from a discretized triangular distribution with parameters \( (w_{\min}\), \( w_\text{mode}\), \( w_{\max})\).
	\item Set the number of groups to \( k=\left\lceil n\cdot (1-r_\text{conc}) \right \rceil \).
	\item Assign one item to each group.
	\item For each remaining item, identify the groups that would not exceed the total group weight of \( 100m\) if this item is assigned to that group. Pick any of those groups randomly with equal probability and assign the item to that group. If there is no such group, then create a new group and assign the item to it.
\end{itemize}

We designed \( r_\text{load}\) and \(r_\text{conc}\) such that \( r_\text{load} \sim \text{Uniform}(1,20)\) and \(r_\text{conc}\sim \text{Uniform}(0,1)\).
The instance generation method can slightly exceed the desired \( r_\text{load}\)
and can have a smaller \(r_\text{conc}\);
so we tested the distributions of both ratios with a one-sample Kolmogorov-Smirnov test \citep{massey1951kolmogorov}.
In both cases the \(p\)-values are over 0.995, therefore the null hypothesis that both ratios have the desired uniform distributions is not rejected.

\subsection{Results for bi-GMKP Algorithms}

\begin{figure}[ht]\caption{Maximum exceeded knapsack capacity per bi-GMKP algorithm}\label{fig:exceeded_cap0}
	\begin{tabularx}{\linewidth}{XXXXXX}
		 & (a) &  & (b) &  & (c)
	\end{tabularx}
	\includegraphics[width=\linewidth]{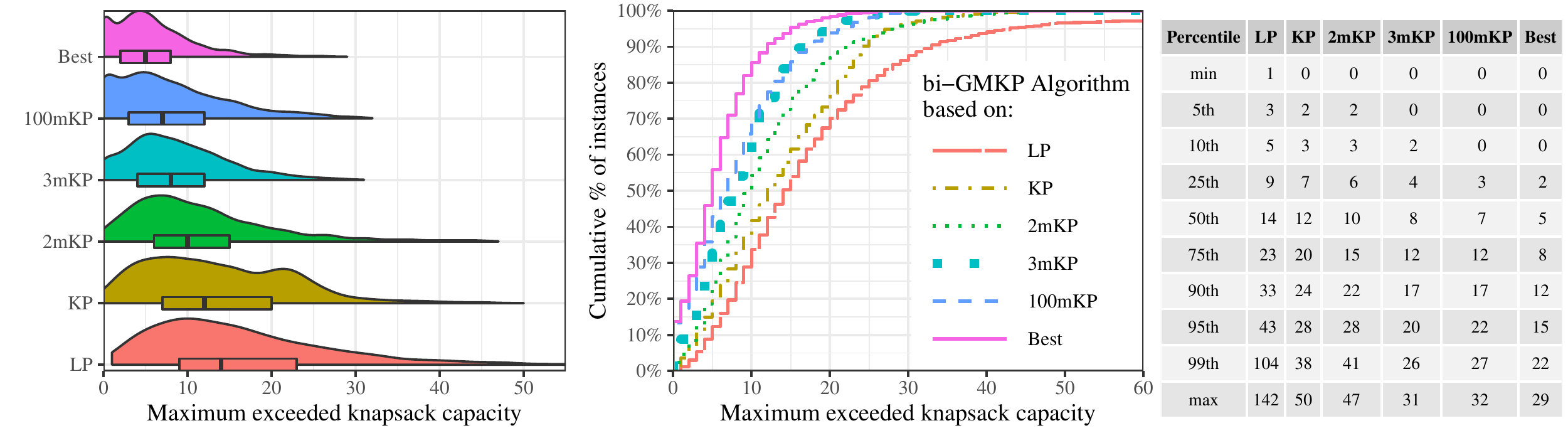}
\end{figure}

\begin{figure}[ht]\caption{Maximum exceeded knapsack capacity per bi-GMKP algorithm after swap-optimal improvement}\label{fig:exceeded_cap2}
	\begin{tabularx}{\linewidth}{XXXXXX}
		 & (a) &  & (b) &  & (c)
	\end{tabularx}
	\includegraphics[width=\linewidth]{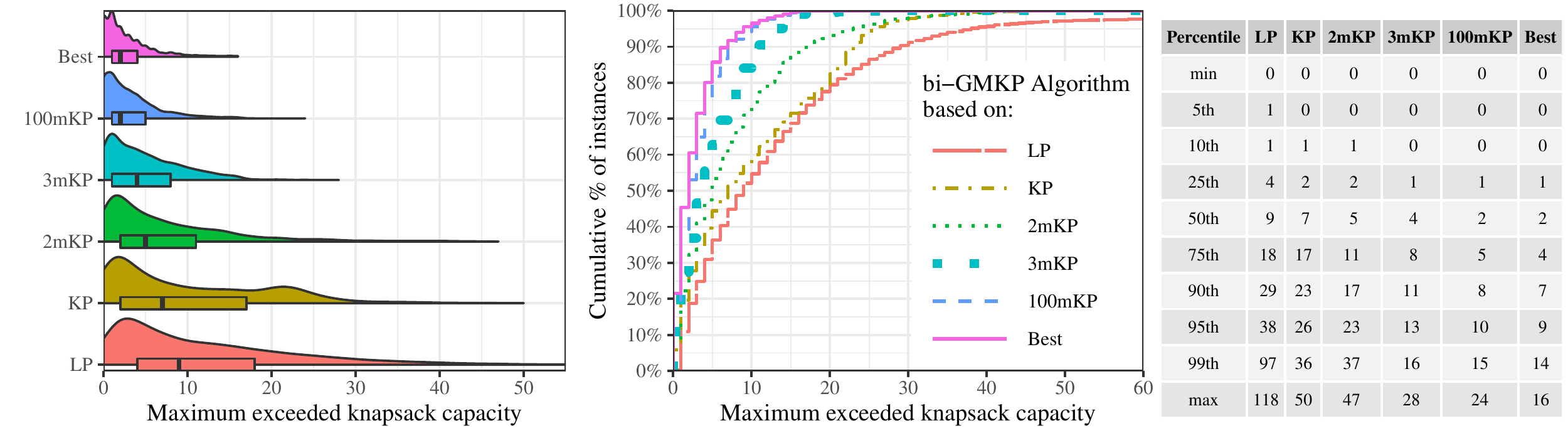}
\end{figure}

Plots (a) in \Cref{fig:exceeded_cap0,fig:exceeded_cap2} show a box plot of the maximum exceeded knapsack capacities obtained by bi-GMKP algorithms on all instances, without and with swap-optimal improvement, respectively.
The box plots mark the 25th, 50th, and 75th percentiles and contain a density plot.
Plots (b) of the figures show the respective cumulative density plots of each algorithm, and (c) show a summary table containing some percentile values of the maximum exceeded knapsack capacity.
Note the complexity and solution quality of each bi-GMKP algorithm increases from LP to 100mKP as ordered in the figures.
Results improve significantly after doing a swap-optimal improvement, where 3mKP exceeds the capacity by at most 16 in 99\% of instances.

\begin{figure}[ht]\caption{Computation time per bi-GMKP algorithm after swap-optimal improvement}\label{fig:comp_time}
	\begin{tabularx}{\linewidth}{XXXX}
		 & (a) &  & (b) 
	\end{tabularx}
	\includegraphics[width=\linewidth]{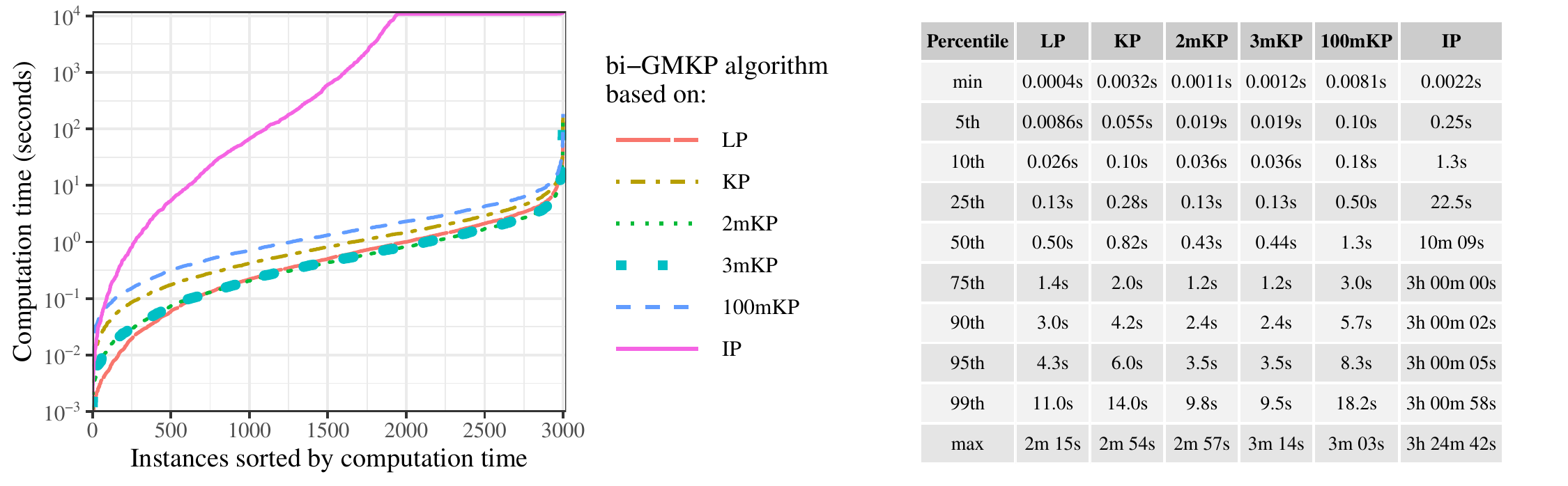}
\end{figure}

Plot (a) of \Cref{fig:comp_time} shows a logarithmic graph of the computation times of each bi-GMKP algorithm after swap-optimal improvement (times are sorted); (b) shows a summary table containing some percentile values of the comptation times.
Over 33\% of instances were not solved by IP in three hours, while each bi-GMKP algorithm ran for less than 4 minutes on each instance; 99\% of instances were solved by each algorithm in less than 19 seconds.
The 3mKP bi-GMKP algorithm achieves a good balance between computation times and  performance (see \Cref{fig:exceeded_cap0,fig:exceeded_cap2}).
Adding some constraints seems to improve computation time, but including too many constraints slows down bi-GMKP algorithm.
Although, using many constraints is still very fast and obtained the best results.

Whenever maximizing rewards is a priority and we can slightly exceed capacities, we recommend using 3mKP because it balances performance and computation time, and has the best approximation guarantee of \(\beta \le 1/3\).
In most cases, 3mKP runs faster than the LP, KP, and 2mKP variations of bi-GMKP algorithm, while having a lower maximum exceeded knapsack capacity.
We also recommend using 100mKP when computation time is not an issue, since 100mKP obtains slightly better results than 3mKP in most instances.
Both 3mKP and 100mKP solve most instances in less than 20 seconds, so using 100mKP (or other variation with many additional constraints) is encouraged.

\subsection{Results for GMKP Heuristics}\label{ssec:results_GMKP_heur}

All bi-GMKP algorithms, after swap-optimal improvement, were also tested with the binary-search GMKP heuristic (\hyperref[heur:binary-GMKP2]{Heuristic~\ref*{heur:binary-GMKP}}) to find feasible GMKP solutions.
The optimal reward ratio corresponds to the reward obtained by the respective GMKP heuristic, divided by the reward of GMKP's optimal solution.
Since we could not find the optimal solution of some instances, in such cases we re-ran the IP solver for up to three more hours; warm starting it from the best feasible solution found so far.
If after such run Gurobi could not determine an optimal solution, we instead set the denominator of the optimal reward ratio to be the best reward found increased by the duality gap.
Note 95\% of all 3,000 instances had a gap smaller than 4.3\%, and 99\% of instances a gap smaller than 10.7\%.

\begin{figure}[ht]\caption{Optimal reward ratio per GMKP heuristic after swap-optimal improvement}\label{fig:optimal_heur}
	\begin{tabularx}{\linewidth}{XXXXXX}
		 & (a) &  & (b) &  & (c)
	\end{tabularx}
	\includegraphics[width=\linewidth]{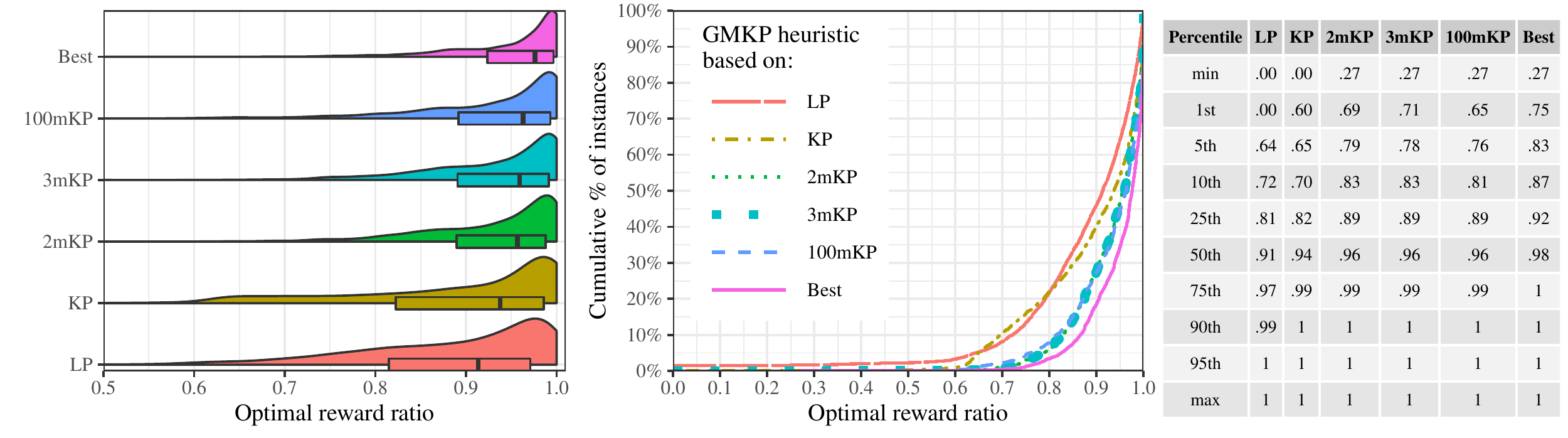}
\end{figure}

Results of the optimal reward ratio appear in \Cref{fig:optimal_heur};
(a) shows the box plots and density plots of the optimal reward ratio obtained by GMKP heuristics on all instances,
(b) shows the respective cumulative density plots of each heuristic,
and (c) shows a summary table containing some percentile values of the optimal reward ratio.
No significant improvement is obtained when having additional constraints beyond 2mKP\@.
Only 5\% of instances solved by 2mKP had an optimal reward ratio worse than 0.79, and only 1\% worse than 0.69.

\begin{figure}[ht]\caption{Computation time per GMKP heuristic after swap-optimal improvement}\label{fig:comp_time2}
	\begin{tabularx}{\linewidth}{XXXX}
		 & (a) &  & (b) 
	\end{tabularx}
	\includegraphics[width=\linewidth]{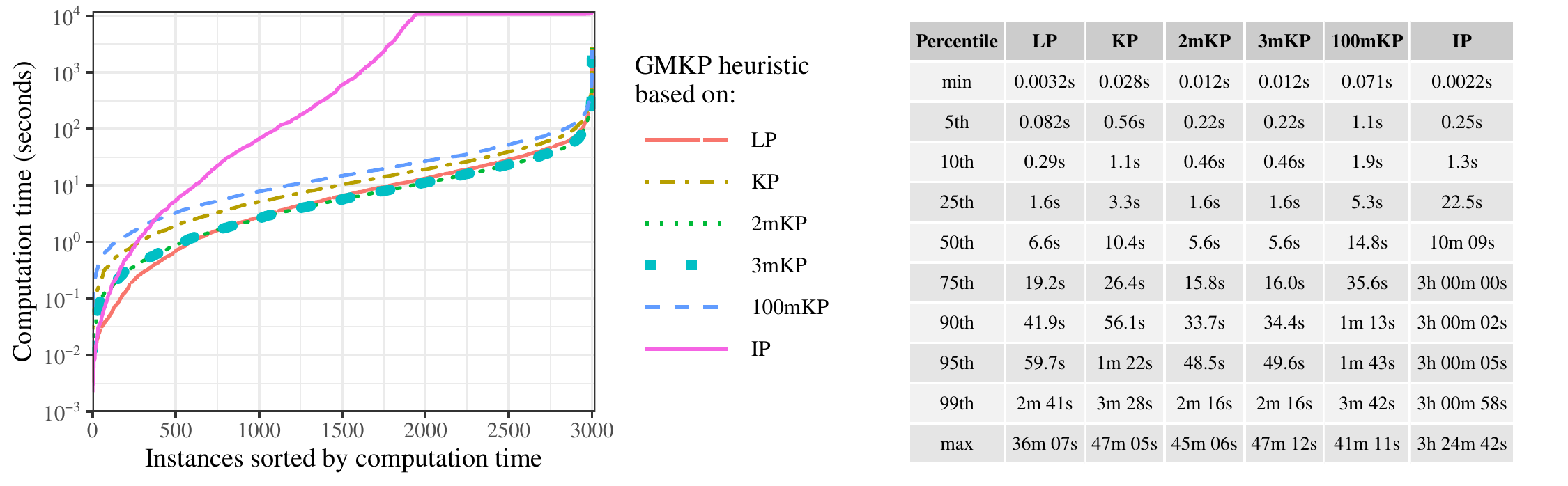}
\end{figure}

\Cref{fig:comp_time2} shows the computation times of GMKP heuristics;
Plot (a) shows a logarithmic graph of the computation time of each GMKP heuristic after swap-optimal improvement; (b) shows a summary table containing some percentile values of the computation times.
Every heuristic ran on any instance in less than 48 minutes, and on 99\% of instances in less than 4 minutes.
Although some instances might take longer, the binary-search GMKP heuristic can stop at any time and return the best feasible solution found.

We recommend using 2mKP based GMKP heuristic when solving GMKP instances, since there is no noticeable improvement when adding additional constraints (such as 3mKP and 100mKP), and because it has the shortest computation time in most instances.
2mKP does not find the optimal solution of most instances, but most of the times it runs in less than 2 minutes obtaining a reward no more than 20\% away from optimal.

\subsection{Results for bi-GMKP Heuristics}
We also tested a bi-GMKP heuristic by  using \hyperref[heur:bi-GMKP2]{Heuristic~\ref*{heur:bi-GMKP}} with the
modified versions of \Cref{alg:2mKP-GMKP} (2mKP).
We tested 2mKP since the results of GMKP heuristics do not improve after adding additional constraints (see \Cref{fig:optimal_heur} of \Cref{ssec:results_GMKP_heur}).
In order to generate a non-dominated frontier of instances, we ran \hyperref[heur:bi-GMKP2]{Heuristic~\ref*{heur:bi-GMKP}} for several \textit{TotalCapacity} values;
multiplying each instance's total capacity by factors going from 0.75 to 1.25 (with 0.05 increments; the heuristic ran 11 times per instance).

\begin{figure}[ht]\caption{(a) Non-dominated frontier example \& (b) Non-dominated solutions of all instances}\label{fig:Pareto_example}
	\begin{tabularx}{\linewidth}{XXXX}
		 & (a) &  & (b) 
	\end{tabularx}
	\includegraphics[width=0.48\textwidth]{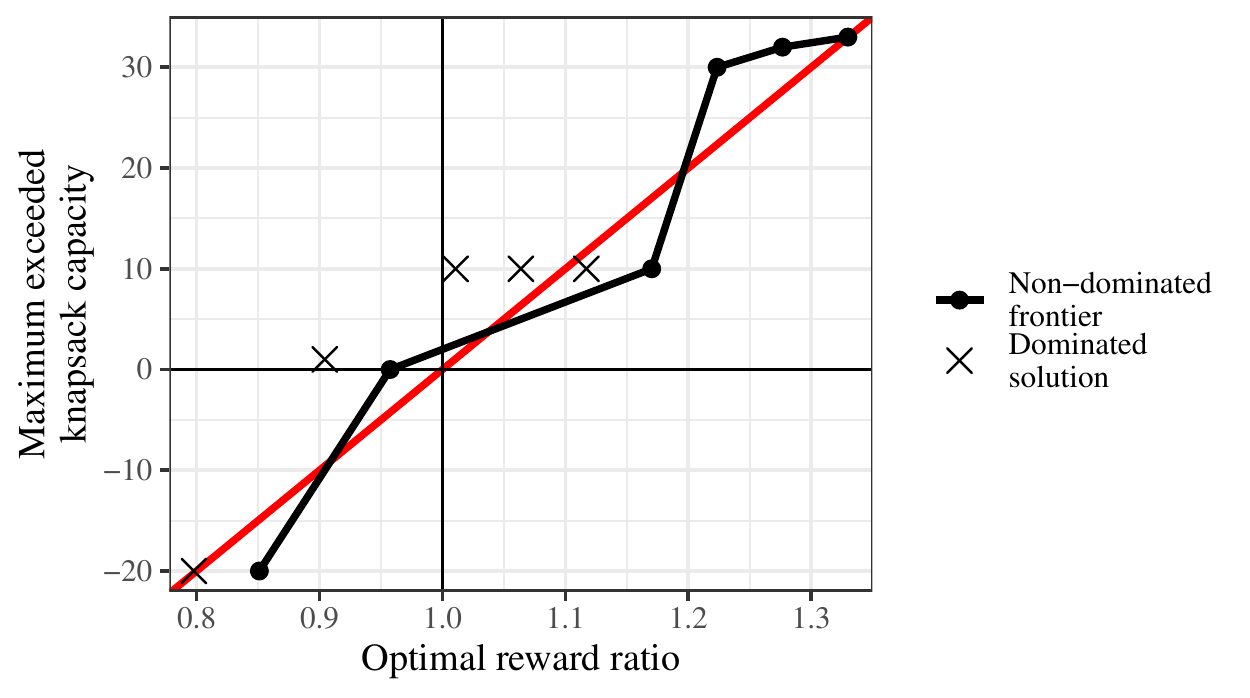}
	\includegraphics[width=0.48\textwidth]{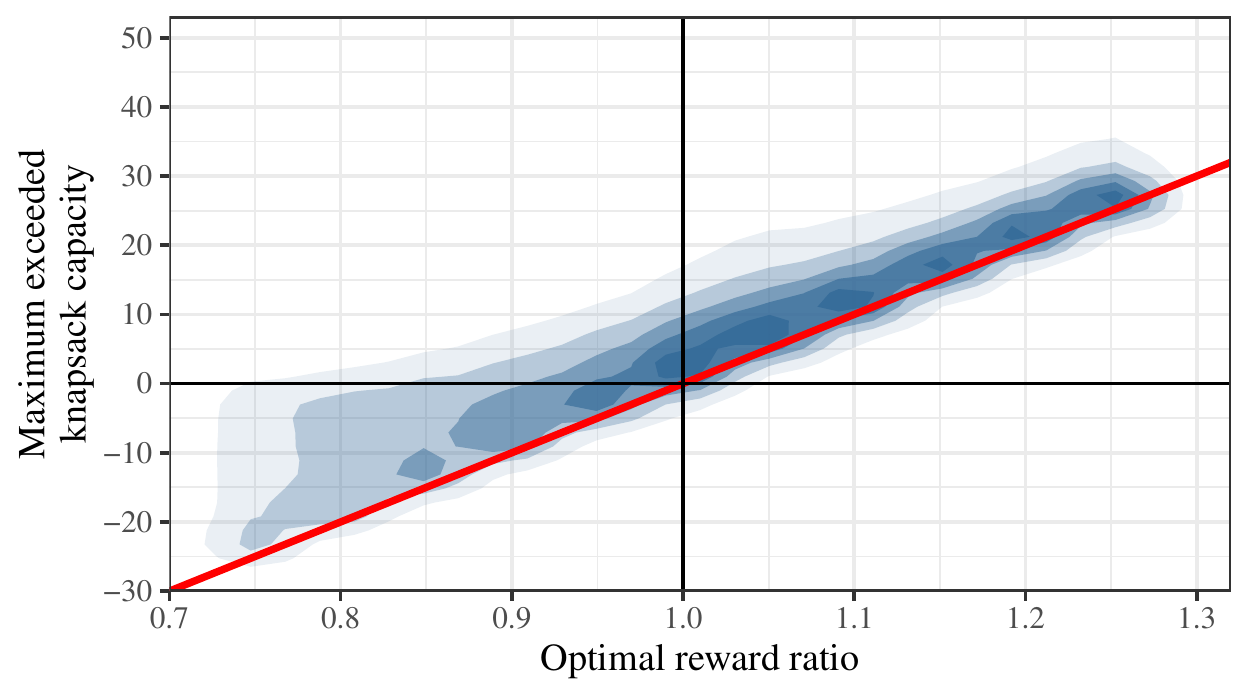}
\end{figure}

Plot (a) of \Cref{fig:Pareto_example} shows an example of a single bi-GMKP instance with several solutions obtained by 2mKP based \hyperref[heur:bi-GMKP2]{Heuristic~\ref*{heur:bi-GMKP}}, for different \textit{TotalCapacity} values.
Each solution has a combination of optimal reward ratio and maximum exceeded knapsack capacity, and some are dominated by other solutions
(i.e., the dominated solution does worse than another solution in both objectives).
Note that negative values of the maximum exceeded knapsack capacity can be obtained when no capacity is exceeded, and the negative value represents the free capacity of the knapsack with the least free capacity. (E.g., a value of -20 means that the knapsack with least free capacity has 20 unutilized capacity, and therefore all other knapsacks have 20 unutilized capacity or more.) 
The red line represents the case where changes in the optimal reward ratio generate a proportional change in the maximum exceeded knapsack capacity.

Plot (b) of \Cref{fig:Pareto_example} shows the contour plot of the non-dominated solutions of all 3,000 instances; where darker means that more solutions are in the area.
Note 11.3\% of the 33,000 solutions were dominated, meaning that the contour plot (b)  includes 88.7\% of the solutions obtained.
Most solutions lie above the red line, showing that the heuristic does a good job in maximizing rewards while slightly exceeding capacities.
Therefore, \hyperref[heur:bi-GMKP2]{Heuristic~\ref*{heur:bi-GMKP}} can be used to obtain different bi-GMKP solutions to decide among different bi-criteria combinations.

We recommend using \hyperref[heur:bi-GMKP2]{Heuristic~\ref*{heur:bi-GMKP}} in practice, since it generates several alternative solutions to pick from, that allow decision makers to evaluate the trade-off between exceeding knapsack capacities and maximizing rewards.
\hyperref[heur:bi-GMKP2]{Heuristic~\ref*{heur:bi-GMKP}} generates solutions that efficiently maximize rewards while barely exceeding knapsack capacities.

\section{Concluding Remarks}\label{SEC:conclusion}

This paper studies GMKP, a strongly \( \mathcal{NP}\)-hard problem with no polynomial time approximation algorithm.
We are not the first studying GMKP \citep{chen2018packing}, but we are the first studying the bi-criteria version of GMKP and offering a broad computational study of our suggested algorithms and heuristics.
We propose several pseudo-polynomial time approximation algorithms for bi-GMKP with tight guarantees, that can be adapted as binary-search heuristics for GMKP\@, and heuristics for bi-GMKP\@.

The proposed algorithms for bi-GMKP either solve a KP or an mKP (KP with multiple capacity constraints) to maximize rewards, and then assign all picked items among knapsacks to minimize the maximum exceeded knapsack capacity.
bi-GMKP algorithms can be combined with binary-search heuristics to obtain capacity feasible GMKP solutions.
In a similar way, modified versions of bi-GMKP algorithms can also be used to obtain different bi-criteria combinations of rewards and maximum exceeded knapsack capacities.

An extensive computational study shows that algorithms and heuristics run fast and obtain good results.
We tested a total of 3,000 instances that had 2 to 100 knapsacks, where the total item weight can be 1 to 20 times larger than the total knapsack capacities.
In 99\% of instances, approximation algorithms for bi-GMKP ran in less than 19 seconds and exceeded the knapsack capacities by at most 16\%.
In 95\% of the cases, GMKP heuristics obtained feasible solutions in less than a minute where the worst reward obtained was only 21\% below the optimal reward; 75\% of the cases, only 11\% below the optimal reward.
Running a swap-optimal improvement after running any algorithm/heuristic is greatly encouraged since they run in polynomial time and improve solutions significantly.

Patient scheduling is one application of bi-GMKP, where the suggested algorithms and heuristics give the decision maker a good set of tools to analyze the trade-off between rewards and maximum exceeded knapsack capacities.

\bibliographystyle{plainnat}
\bibliography{refs.bib}

\begin{appendices}
	% remove some chktex warnings
% chktex-file 1 % cmd terminated with space
% chktex-file 9 % ) expected, found } 
% chktex-file 21 % "this command might not be intended" \left\{ bug
% chktex-file 10 % "10: Solo `]' found." bug for \label[... in appendix

\section{Greedy LP-GMKP Algorithm}\label[appendix]{APPENDIX:LP-GMKP}

\begin{proposition}\label{prop:extreme_points}
	Optimal extreme points of an LP-GMKP instance can have more than one partially assigned group.
\end{proposition}
\proof{Proof of \Cref{prop:extreme_points}}:
Consider the case with two knapsacks of capacities \({c_1=3}\) and \({c_2=1}\), and two groups with rewards \(p_1=p_2=3\).
The first group has two items that weigh \(w_{1_a}=1\) and \(w_{1_b}=2\), and the second group has one item with \(w_{2}=3\).
Consider the solution \((x,z)\) where:
\begin{align*}
	z & =(z_1,z_2)=\left(1/2,5/6\right)
	\\
	x & =(x_{11_a},x_{11_b},x_{12},x_{21_a},x_{21_b},x_{22})
	=\left(1/2,0,5/6,0,1/2,0\right)
\end{align*}

Solution \((x,z)\) is feasible and optimal, with two partially assigned groups.
Also, it is an extreme point since it has 8 variables and 8 activate linearly independent constraints that are:
\begin{align*}
	3        & =x_{11_a}+2x_{11_b}+3x_{12},          & \text{(from constraints~\eqref{eq:IP-GMKP-cap})} \\
	1        & =x_{21_a}+2x_{21_b}+3x_{22}           & \text{(from constraints~\eqref{eq:IP-GMKP-cap})} \\
	z_1      & =x_{11_a}+x_{21_a}=x_{11_b}+x_{21_b}, & \text{(from constraints~\eqref{eq:IP-GMKP-eq})}  \\
	z_2      & =x_{12}+x_{22}                        & \text{(from constraints~\eqref{eq:IP-GMKP-eq})}  \\
	x_{11_b} & =x_{21_a}=x_{22}=0,                   & \text{(from constraints \(x_{ij}\ge 0\))}
\end{align*}
\Halmos{}
\endproof

\begin{algorithm}[ht]\DontPrintSemicolon
	\caption{Greedy LP-GMKP}\label{alg:greedyLP}
	\KwData{LP-GMKP instance.}
	\KwResult{\( x_{ij}\in[0,1]\), \( \forall i\in M,j\in N\); \( z_l\in[0,1]\), \( \forall l\in K\). }
	Enumerate groups in \( K\)  such that
	\( \frac{p_1}{\sum\limits_{j\in G_1}w_j}\ge\frac{p_2}{\sum\limits_{j\in G_2}w_j}\ge\dots\ge \frac{p_k}{\sum\limits_{j\in G_k}w_j}\) \;\label{alg-line:LP_line_sort}
	\textbf{Initialize:}
	\( x_{ij} \gets 0\), \( \forall i\in M, j\in N\);
	\( z_l \gets 0\), \( \forall l\in K\);
	\( l' \gets 1\);
	\( i' \gets 1\);
	
	\nonl\( KnapsackWeight \gets 0\);
	\( TotalWeight \gets 0\) \;
	
	\While{\( l'\le k\)  \textbf{\upshape and} \( i'\le m\) }{\label{alg-line:LP_line_while}
	\( z_{l'}\gets \min\left(					1,
	\dfrac{\sum\limits_{i\in M}c_i-TotalWeight}{\sum\limits_{j\in G_{l'}}w_j}
	\right)
	\) \;\label{alg-line:LP_z}
	\ForEach{ \( j'\in G_{l'}\)  }{\label{alg-line:LP_line_foreach}
	\( ItemWeight \gets  z_{l'}w_{j'}\) \;
	\While{\( ItemWeight > 0\) }{\label{alg-line:LP_line_while2}
	\( x_{i'j'} \gets
	\dfrac{\min\left(ItemWeight,  c_{i'}-KnapsackWeight
		\right)}{w_{j'}}\) \;\label{alg-line:LP_x}
	\( ItemWeight \gets ItemWeight-x_{i'j'}w_{j'}\) \;
	\( KnapsackWeight\gets KnapsackWeight+x_{i'j'} w_{j'}\) \;
	\( TotalWeight\gets TotalWeight+x_{i'j'} w_{j'}\) \;
	\If{\( KnapsackWeight=c_{i'}\) }{
	\( KnapsackWeight\gets 0\) \;
	\( i' \gets i'+1\) \;
	}
	}
	}
	\( l' \gets l'+1\) \;
	}
\end{algorithm}

\begin{proposition}\label{prop:LP-GMKP}
	\Cref{alg:greedyLP} {\normalfont (i)}   generates an optimal solution for any feasible LP-GMKP instance, and {\normalfont (ii)} runs in polynomial time.
\end{proposition}
\proof{Proof of \Cref{prop:LP-GMKP}}:
(i)
The algorithm  sorts all groups in a non-increasing reward to total weight ratio and then greedily assigns their items into the knapsacks, filling them one by one.
Hence, there are no other groups that can fill the knapsacks with higher rewards; i.e., the solution, if feasible, would be optimal.

To guarantee feasibility, the algorithm checks before assigning a new group \( l\in K\)  if it fits into the remaining capacity (considering all knapsacks).
If it does, it assigns the group (\( z_l=1\), \( {\sum_{i\in M}x_{ij}=1}\), \( {\forall j\in G_l}\)).
If the group does not fit entirely, it will fill up all the remaining capacity (\(z_l<1\) and \( {\sum_{i\in M}x_{ij}=z_l}\), \( {\forall j\in G_l}\)).
In both cases, constraints~\eqref{eq:IP-GMKP-eq} are satisfied.

Finally, capacity constraints~\eqref{eq:IP-GMKP-cap} are satisfied because the knapsacks are filled up one by one, and whenever an item does not fit the current knapsack, it is fractionally split among the current knapsack and the next.
Thus, the solution is feasible.

(ii)
The sorting (\cref{alg-line:LP_line_sort}) runs in polynomial time \( \mathcal{O}(k\log k)\).
The first \enquote{while} loop (\cref{alg-line:LP_line_while}) iterates no more than \( k\)  times,
the \enquote{for each} loop (\cref{alg-line:LP_line_foreach}) iterates no more than \( n\)  times,
and the second \enquote{while} loop (\cref{alg-line:LP_line_while2}) iterates no more than \( m\)  times.
Combining all loops together, in the worst case scenario, the algorithm iterates over \( k\)  groups to go through all \( n\)  items, while also iterating through all \( m\)  knapsacks.
Thus the algorithm runs in polynomial time \( {\mathcal{O}(k\log k+n+m)}\).
\Halmos
\endproof

\begin{corollary}\label{cor:part_groups}
	Any solution found by \Cref{alg:greedyLP} {\normalfont}	has at most one partially assigned group.
\end{corollary}
\proof{Proof of \Cref{cor:part_groups}}:
By construction, only the last group assigned can have \( {0<z_{l}<1}\).\Halmos{}
\endproof

\section{Generalized mKP Based Pseudo-Polynomial Time Approximation Algorithm for bi-GMKP}\label[appendix]{APPENDIX:mKPD}

\begin{definition}
	Given a GMKP instance and a finite set \( {D\subset\mathbb{R}_{>0}}\):
	\begin{align}
		\textbf{mKP}_{\pmb D}\textbf{-GMKP:\qquad}\max & \sum_{l\in K} p_l z_l \tag{\ref*{eq:IP-GMKP-obj}}                                                                     \\
		\text{s.t.}                                    & \sum_{l\in K}\sum_{j\in G_l} w_j z_l\le \sum_{i\in M} c_i\tag{\ref*{eq:KP-GMKP-cap}}\label{eq:KP-GMKP-cap3}           \\
		                                               & \sum_{l\in K}\sum_{j\in G_l} f_d(w_j) z_l\le \sum_{i\in M} f_d(c_i)                                         & d\in D
		\label{eq:mKPD-GMKP-constr}                                                                                                                                            \\
		                                               & z_l\in \{ 0, 1 \}                                                                                           & l \in K
		\nonumber
	\end{align}
\end{definition}
\(\text{mKP}_\text{D}\)-GMKP generalizes KP-GMKP, 2mKP-GMKP and 3mKP-GMKP\@.
KP-GMKP corresponds to \(\text{mKP}_\text{D}\)-GMKP for \(D=\emptyset\),
2mKP-GMKP corresponds to \(\text{mKP}_\text{D}\)-GMKP for \(D=\left\{ c_{\max}/2 \right\}\), and 
3mKP-GMKP to \(\text{mKP}_\text{D}\)-GMKP for \(D=\left\{ c_{\max}/2,c_{\max}/3 \right\}\).

\begin{proposition}\label{prop:minset_D}
	Let \( Z(D)\)  be the feasible region of an \( \text{mKP}_\text{D}\)-GMKP instance, and let
	\begin{align}\label{eq:D}
		D'=(0,w_{\max})\cap \left \{ \frac{c_i}{q}:i\in M, q\in \mathbb{Z}_{>0}\right \}.
	\end{align}
	Then \( Z(D')= Z(\mathbb{R}_{>0})\).
\end{proposition}
\proof{Proof of \Cref{prop:minset_D}}:
This proposition shows that only finite sets \(D\subset D'\) are worth considering when defining an \( \text{mKP}_\text{D}\)-GMKP instance.

We first prove upper bound \( w_{\max}\).
Whenever \( d\ge w_{\max}\) then \( f_d(w_j)=0\),  \( \forall j\in N\).
This makes the \textit{left-hand side} (lhs) coefficients of constraints~\eqref{eq:mKPD-GMKP-constr} to be 0.

Sort elements of \( D'\)  increasingly and let \( d'>0\)  be such that \( d_h<d'<d_{h+1}\), for some consecutive \( d_h,d_{h+1}\in D'\).
We claim that constraint~\eqref{eq:mKPD-GMKP-constr} for such \( d'\)  is redundant with constraint~\eqref{eq:mKPD-GMKP-constr} for \( d_h\),
i.e.,
\begin{equation}\label{eq:comp_denom_set1}
	\sum_{l\in K}\sum_{j\in G_l}f_{d_h}(w_j)z_l\le \sum_{i\in M} f_{d_h}(c_i)
	\implies
	\sum_{l\in K}\sum_{j\in G_l}f_{d'}(w_j)z_l\le \sum_{i\in M} f_{d'}(c_i)
\end{equation}
If~\eqref{eq:comp_denom_set1} holds, then all such \( d'\)  can be removed from \( D'\), and  \( Z(D')= Z(\mathbb{R}_{>0})\) still holds.
Since \( d_h<d'\), when comparing the lhs of constraints~\eqref{eq:mKPD-GMKP-constr} for \( d'\)  and \( d_h\) we have
\begin{equation}\label{eq:comp_denom_set2}
	\sum_{l\in K}\sum_{j\in G_l}f_{d'}(w_j)z_l
	\le
	\sum_{l\in K}\sum_{j\in G_l}f_{d_h}(w_j)z_l
\end{equation}
As the value of \( d\) increases, the \textit{right-hand side} (rhs) of constraints~\eqref{eq:mKPD-GMKP-constr} only change at the points contained in \( D'\)  by integer amounts. Therefore
\begin{equation}\label{eq:comp_denom_set3}
	\sum_{i\in M} f_{d_h}(c_i)
	=
	\sum_{i\in M} f_{d'}(c_i)
\end{equation}
Combining constraint~\eqref{eq:mKPD-GMKP-constr} for \( d_h\), with~\eqref{eq:comp_denom_set2} and~\eqref{eq:comp_denom_set3}, we get
\[
	\sum_{l\in K}\sum_{j\in G_l}f_{d'}(w_j)z_l
	\le
	\sum_{l\in K}\sum_{j\in G_l}f_{d_h}(w_j)z_l\le \sum_{i\in M} f_{d_h}(c_i)
	=
	\sum_{i\in M} f_{d'}(c_i).
\]
Thus,~\eqref{eq:comp_denom_set1} holds.
\Halmos
\endproof

\begin{algorithm}[ht]\DontPrintSemicolon
	\caption{Generalized mKP based approximation algorithm for bi-GMKP}\label{alg:mKPD-GMKP}
	\KwData{bi-GMKP instance, finite set \( D\subset D'\) as defined in~\eqref{eq:D}.}
	\KwResult{\( x^a_{ij}\in \{ 0, 1 \} \), \( \forall i\in M,j\in N\);
	\( z^a_l\in \{ 0, 1 \} \), \( \forall l\in K\).}
	\nonl Run \Cref{alg:LP-GMKP}, changing \cref{alg-line:LP-GMKP_LP} with the following:
	\\\nonl Solve the corresponding \(\text{mKP}_\text{D}\)-GMKP instance, and get solution \( z^a\).\;
\end{algorithm}

\begin{theorem}\label{theo:approx_mKP-GMKP2}
	Let \( D'\)  be defined as~\eqref{eq:D} (in \Cref{prop:minset_D}),
	and let  \( D\subset D'\)  be a finite set containing \( c_{\max}/2\).
	For such \( D\),
	\Cref{alg:mKPD-GMKP}
	{\normalfont (i)}
	is a \( \left(1,1/2\right)\)-approximation algorithm, and
		{\normalfont (ii)} runs in pseudo-polynomial time.
		{\normalfont (iii)} This is a tight approximation.
\end{theorem}

\proof{Proof of \Cref{theo:approx_mKP-GMKP2}}:
This theorem shows that even when set \( D\)  is very large, the worst case approximation obtained by \( D=\{c_{\max}/2\} \)  is not improved (equivalent to the worst case approximation of \Cref{alg:2mKP-GMKP}).

(i,ii) Analogous to \Cref{theo:approx_mKP-GMKP} proof, since \(\{c_{\max}/2\}\in D\).
The pseudo-polynomial time is
\[
	{\mathcal{O}\left(m\log(m) + n\left(\sum_{i\in M} c_i \right)\prod_{d\in D}\left(\sum_{i\in M} f_{d}(c_i) \right)\right)}.
\]
(iii)
See \hyperref[eg:2mKPD]{Examples for  \Cref{theo:approx_mKP-GMKP2}} in \Cref{APPENDIX:tight}.
\Halmos
\endproof

\begin{corollary}\label{cor:bi-GMKP2}
	Even if \Cref{alg:mKPD-GMKP} could solve an instance for \( D=\mathbb{R}_{>0}\), the \( \left(1,1/2\right)\)-approximation guarantee from \Cref{theo:approx_mKP-GMKP2} does not improve.
\end{corollary}
\proof{Proof of \Cref{cor:bi-GMKP2}}:
Consider the tight example of \Cref{theo:approx_mKP-GMKP2} (\Cref{fig:mKPD-eg} in \Cref{APPENDIX:tight}).
The groups picked by the algorithm have a feasible assignment in the corresponding GMKP instance (by rearranging items).
Since all constraints~\eqref{eq:mKPD-GMKP-constr} are valid inequalities for GMKP (proof analogous to \Cref{lemma:2mKP-GMKP_relax}), then the solution found by the algorithm is not removed by constraints~\eqref{eq:mKPD-GMKP-constr} for any \(d\in D\); thus the tight \(\beta\le 1/2\) example works for \( D=\mathbb{R}_{>0}\). 
\Halmos
\endproof

\begin{theorem}\label{theo:approx_mKP-GMKP3}
	Let \( D'\)  be defined as~\eqref{eq:D} (in \Cref{prop:minset_D}),
	and let  \( D\subset D'\)  be a finite set such that \( \left\{c_{\max}/2,c_{\max}/3 \right\}\subseteq D\).
	For such \( D\), when all knapsacks have equal capacities,
	\Cref{alg:mKPD-GMKP}
	{\normalfont (i)}
	is a \( \left(1,1/3\right)\)-approximation algorithm, and
		{\normalfont (ii)} runs in pseudo-polynomial time.
		{\normalfont (iii)} This is a tight approximation.
\end{theorem}

\proof{Proof of \Cref{theo:approx_mKP-GMKP3}}:
This theorem shows that even when set \( D\)  is very large, the worst case approximation obtained by \( D=\left\{c_{\max}/2,c_{\max}/3\right\} \) (equivalent to the worst case approximation of \Cref{alg:3mKP_GMKP}) is not improved when all knapsacks have equal capacities.

(i,ii) Analogous to \Cref{{theo:approx_3mKP}} proof, since \(\left\{c_{\max}/2,c_{\max}/3\right\}\in D\).
The pseudo-polynomial time is the same as in \Cref{theo:approx_mKP-GMKP2}.

(iii)
See \hyperref[eg:3mKPD]{Examples for  \Cref{theo:approx_mKP-GMKP3}} in \Cref{APPENDIX:tight}.
\Halmos
\endproof

\section{Tight Examples}\label[appendix]{APPENDIX:tight}

\paragraph{Examples for  \Cref{theo:approx_LP-GMKP}.}\label{eg:LP}
The tightness of guarantee \( \alpha\ge 1\) is trivial; any example where the algorithm gives an optimal solution to GMKP works.
\begin{figure}[ht]\caption{Tight \(\pmb{\beta\le 2}\) example for  \Cref{theo:approx_LP-GMKP}}\label{fig:LP-eg}

	\begin{tikzpicture}[x=0.75pt,y=0.75pt,yscale=-1,xscale=1]

		\begin{scope}	
			\node [anchor=west] at (60,5) {LP-GMKP optimal solution};			
			\draw (43,70) node   {$0$};
			\draw (43,20) node   {$1$};
			\draw    (60,20) -- (50,20) ;
			\draw    (60,20) -- (60,70) ;
			\draw    (60,70) -- (50,70) ;
			
			\draw  [fill=lightgray] (80,20) -- (120,20) -- (120,70) -- (80,70) -- cycle ;
			\draw  [fill=lightgray] (130,20) -- (170,20) -- (170,70) -- (130,70) -- cycle ;
			\draw  [fill=lightgray] (210,20) -- (250,20) -- (250,70) -- (210,70) -- cycle ;
			\draw  [fill=lightgray] (260,30) -- (300,30) -- (300,70) -- (260,70) -- cycle ;
			\draw  [pattern=linedownleftright] (260,20) -- (300,20) -- (300,30) -- (260,30) -- cycle ;
			
			\draw (100,45) node [scale=1]  {$1$};
			\draw (150,45) node [scale=1]  {$1$};
			\draw (190.5,44) node   {$\cdots $};
			\draw (231,46) node [scale=1]  {$1$};
			\draw (281,50) node [scale=1]  {$\frac{m}{m+1}$};
		\end{scope}

		\begin{scope}	
			\node [anchor=west] at (375,5) {GMKP optimal solution};	
			\draw (357,70) node   {$0$};
			\draw (357,20) node   {$1$};
			\draw    (374,20) -- (374,70) ;
			\draw    (374,70) -- (364,70) ;
			\draw    (374,20) -- (364,20) ;

			\draw  [color=gray] (569.5,20) -- (609.5,20) -- (609.5,70) -- (569.5,70) -- cycle ;
			\draw  [fill=lightgray] (389.5,20) -- (429.5,20) -- (429.5,70) -- (389.5,70) -- cycle ;
			\draw  [fill=lightgray] (439.5,20) -- (479.5,20) -- (479.5,70) -- (439.5,70) -- cycle ;
			\draw  [fill=lightgray] (519.5,20) -- (559.5,20) -- (559.5,70) -- (519.5,70) -- cycle ;
			\draw  [fill=lightgray] (569.5,30) -- (609.5,30) -- (609.5,70) -- (569.5,70) -- cycle ;
			
			\draw (409.5,45) node [scale=1]  {$1$};
			\draw (459.5,45) node [scale=1]  {$1$};
			\draw (500,44) node   {$\cdots $};
			\draw (539.5,45) node [scale=1]  {$1$};
			\draw (589.5,50) node [scale=1]  {$\frac{m}{m+1}$};
		\end{scope}

		\begin{scope}[shift={(0,-10)}]
			\node [anchor=west] at (60,120) {\Cref{alg:LP-GMKP} bi-GMKP solution};
			\draw (43,260) node   {$0$};
			\draw (43,210) node   {$1$};
			\draw (43,160) node   {$2$};
			\draw (43,110) node   {$3$};
			\draw    (60,110) -- (60,260) ;
			\draw    (60,110) -- (50,110) ;
			\draw    (60,160) -- (50,160) ;	
			\draw    (60,210) -- (50,210) ;	
			\draw    (60,260) -- (50,260) ;
			
			\draw  [color=gray] (260,210) -- (300,210) -- (300,260) -- (260,260) -- cycle ;
			\draw  [fill=lightgray] (80,210) -- (120,210) -- (120,260) -- (80,260) -- cycle ;
			\draw  [fill=lightgray] (130,210) -- (170,210) -- (170,260) -- (130,260) -- cycle ;
			\draw  [fill=lightgray] (210,210) -- (250,210) -- (250,260) -- (210,260) -- cycle ;
			\draw  [fill=lightgray] (260,220) -- (300,220) -- (300,260) -- (260,260) -- cycle ;

			\draw[pattern=linedownleftright]   (80,170) -- (120,170) -- (120,210) -- (80,210) -- cycle ;
			\draw[pattern=linedownleftright]   (130,170) -- (170,170) -- (170,210) -- (130,210) -- cycle ;
			\draw[pattern=linedownleftright]   (210,170) -- (250,170) -- (250,210) -- (210,210) -- cycle ;
			\draw[pattern=linedownleftright]   (260,180) -- (300,180) -- (300,220) -- (260,220) -- cycle ;
			\draw[pattern=linedownleftright]   (260,140) -- (300,140) -- (300,180) -- (260,180) -- cycle ;

			\draw (98.5,235) node [scale=1]  {$1$};
			\draw (148.5,235) node [scale=1]  {$1$};
			\draw (190.5,216) node   {$\cdots $};
			\draw (229.5,236) node [scale=1]  {$1$};
			\node at (279.5,240) {$\frac{m}{m+1}$};
			\node at (99.5,190)  {\contour{white}{$\frac{m}{m+1}$}};
			\node at (149.5,190) {\contour{white}{$\frac{m}{m+1}$}};
			\node at (229.5,190) {\contour{white}{$\frac{m}{m+1}$}};
			\node at (279.5,160) {\contour{white}{$\frac{m}{m+1}$}};
			\node at (279.5,195) {\contour{white}{$\frac{m}{m+1}$}};

			\draw   (305,210.14) .. controls (309.67,210.14) and (312,207.81) .. (312,203.14) -- (312,185.43) .. controls (312,178.76) and (314.33,175.43) .. (319,175.43) .. controls (314.33,175.43) and (312,172.1) .. (312,165.43)(312,168.43) -- (312,147.71) .. controls (312,143.04) and (309.67,140.71) .. (305,140.71) ;			
			\draw (355,177) node   {$\frac{3m}{m+1}-1$};
			\draw (365,200) node   {$=2-\frac{3}{m+1}$};
		\end{scope}
		
		\begin{scope}[shift={(-10,-40)}]%legend
			\draw (550,149) node  [align=left] {Items of group 1};
			\draw (550,190) node  [align=left] {Items of group 2};
			\draw (520,232) node  [align=left] {Capacity};
			
			\draw   (450,130) -- (620,130) -- (620,250) -- (450,250) -- cycle ;
			
			\draw  [fill=lightgray] (460,140) -- (480,140) -- (480,160) -- (460,160) -- cycle ;
			\draw  [pattern=linedownleftright] (460,180) -- (480,180) -- (480,200) -- (460,200) -- cycle ;
			\draw[color=gray]    (460,220) -- (480,220) -- (480,240) -- (460,240) -- cycle ;
		\end{scope}

	\end{tikzpicture}
\end{figure}
Refer to \Cref{fig:LP-eg} for the tight \( \beta\le 2\)  example.
Consider \( m\ge 3\)  knapsacks of equal capacities 1,  and two groups where
\begin{itemize}
	\item Group 1 has \( m-1\)  items that weigh 1 each and one item that weighs \(m/(m+1)\).
	\item Group 2 has \( m+1\)  items that weigh \(m/(m+1)\).
\end{itemize}
If all rewards equal total group weights then,
given this ordering of groups,
\Cref{alg:LP-GMKP} generates a solution where \( m-1\) knapsacks each contain two items of weights 1 and \( m/(m+1)\).
The last knapsack contains three items of weight \( m/(m+1)\).
Therefore, the maximum exceeded knapsack capacity is \( {2-3/(m+1)}\),
and as \( m\to \infty \) it converges to 2.

\paragraph{Examples for  \Cref{theo:approx_KP-GMKP}.}\label{eg:KP}
The tightness of guarantee \( \alpha\ge 1\) is trivial; any example where the algorithm gives an optimal solution to GMKP works.
For the tight \( \beta\le 1\) example, refer to the same instance as in \hyperref[eg:LP]{Examples for  \Cref{theo:approx_LP-GMKP}}, but only considering group 2 with \( m+1\)  items that weigh \(m/({m+1})\). 
\Cref{alg:KP-GMKP} generates a solution where \( m-1\)  knapsacks have one item assigned, and one knapsack has two items  assigned.
Therefore, the maximum exceeded knapsack capacity is \( {1-2/(m+1)}\),  and as \( m\to \infty \) the bound converges to \( 1\).
Note constraint~\eqref{eq:KP-GMKP-cap} of KP-GMKP is satisfied.

\paragraph{Examples for  \Cref{theo:approx_mKP-GMKP}.}\label{eg:mKP}
The tightness of guarantee \( \alpha\ge 1\) is trivial; any example where the algorithm gives an optimal solution to GMKP works.
\begin{figure}[ht]\caption{Tight \(\pmb{\beta\le 1/2}\) example for  \Cref{theo:approx_mKP-GMKP}}\label{fig:mKP-eg}
	\begin{tikzpicture}[x=0.75pt,y=0.75pt,yscale=-1,xscale=1]
		
		\begin{scope}
			\node [anchor=west] at (70,65) {GMKP optimal solution};
			\draw (43,133) node   {$0$};
			\draw (43,83) node   {$1$};
			\draw    (60,81) -- (60,131) ;
			\draw    (60,81) -- (50,81) ;
			\draw    (60,131) -- (50,131) ;
			
			\draw (190.5,107) node   {$\cdots $};
			
			\draw  [color=gray] (80,81) -- (120,81) -- (120,131) -- (80,131) -- cycle ;
			\draw  [color=gray] (130,81) -- (170,81) -- (170,131) -- (130,131) -- cycle ;
			\draw  [color=gray] (210,81) -- (250,81) -- (250,131) -- (210,131) -- cycle ;
			\draw  [color=gray] (260,81) -- (300,81) -- (300,131) -- (260,131) -- cycle ;
		\end{scope}
		
		\begin{scope}
			\node [anchor=west] at (70,175) {\Cref{alg:2mKP-GMKP} bi-GMKP solution};	
			\draw (43,260) node   {$0$};
			\draw (43,210) node   {$1$};
			\draw (43,160) node   {$2$};		
			\draw    (60,160) -- (60,260) ;
			\draw    (60,160) -- (50,160) ;
			\draw    (60,210) -- (50,210) ;
			\draw    (60,260) -- (50,260) ;

			\draw  [color=gray] (80,210) -- (120,210) -- (120,260) -- (80,260) -- cycle ;
			\draw  [color=gray] (130,210) -- (170,210) -- (170,260) -- (130,260) -- cycle ;
			\draw  [color=gray] (210,210) -- (250,210) -- (250,260) -- (210,260) -- cycle ;			
			\draw  [fill=lightgray] (260,200) -- (300,200) -- (300,220) -- (260,220) -- cycle ;			
			\draw  [color=gray] (260,210) -- (300,210) -- (260,210) -- (300,210) -- cycle ;
			
			\draw  [fill=lightgray] (80,220) -- (120,220) -- (120,240) -- (80,240) -- cycle ;
			\draw  [fill=lightgray] (80,240) -- (120,240) -- (120,260) -- (80,260) -- cycle ;
			\draw  [fill=lightgray] (130,220) -- (170,220) -- (170,240) -- (130,240) -- cycle ;
			\draw  [fill=lightgray] (130,240) -- (170,240) -- (170,260) -- (130,260) -- cycle ;
			\draw  [fill=lightgray] (210,220) -- (250,220) -- (250,240) -- (210,240) -- cycle ;
			\draw  [fill=lightgray] (210,240) -- (250,240) -- (250,260) -- (210,260) -- cycle ;
			\draw  [fill=lightgray] (260,220) -- (300,220) -- (300,240) -- (260,240) -- cycle ;
			\draw  [fill=lightgray] (260,240) -- (300,240) -- (300,260) -- (260,260) -- cycle ;
			
			\draw (100,230) node [scale=1]  {$\frac{m}{2m+1}$};
			\draw (100,250) node [scale=1]  {$\frac{m}{2m+1}$};
			\draw (150,230) node [scale=1]  {$\frac{m}{2m+1}$};
			\draw (150,250) node [scale=1]  {$\frac{m}{2m+1}$};
			\draw (190.5,235) node   {$\cdots $};
			\draw (230,230) node [scale=1]  {$\frac{m}{2m+1}$};
			\draw (230,250) node [scale=1]  {$\frac{m}{2m+1}$};
			\draw (280,210) node [scale=1]  {$\frac{m}{2m+1}$};
			\draw (280,230) node [scale=1]  {$\frac{m}{2m+1}$};
			\draw (280,250) node [scale=1]  {$\frac{m}{2m+1}$};
			
			\node[scale=0.8] [anchor=west] at (299,205) {$\}$};
			\node [anchor=west] at (305,205) {$\frac{3m}{2m+1}-1=\frac{1}{2}-\frac{3}{4m+2}$};
		\end{scope}
		
		\begin{scope}[shift={(-105,0)}]%legend
			\draw (530,102) node  [align=left] {Items of group 1};
			\draw (500,142) node  [align=left] {Capacity};
			
			\draw  [color=gray] (440,130) -- (460,130) -- (460,150) -- (440,150) -- cycle ;
			\draw  [fill=lightgray] (440,90) -- (460,90) -- (460,110) -- (440,110) -- cycle ;
			
			\draw   (430,80) -- (600,80) -- (600,160) -- (430,160) -- cycle ;
		\end{scope}
		
	\end{tikzpicture}
	
\end{figure}
Refer to \Cref{fig:mKP-eg} for the tight \( \beta\le 1/2\)  example.
Consider \( m\ge 3\)  knapsacks of equal capacities 1,  and one group that has \( 2m+1\)  items that weigh \(m/({2m+1})\) each.
\Cref{alg:2mKP-GMKP} generates a solution where \( m-1\) knapsacks each contain two items and one knapsack contains three items.
Therefore, the maximum exceeded knapsack capacity is \( 1/2-3/({4m+2})\), and as \( m\to \infty \) it converges to \(1/2\).
Note constraints~\eqref{eq:KP-GMKP-cap2} and~\eqref{eq:2mKP-GMKP-constr}  of 2mKP-GMKP are satisfied.

\paragraph{Examples for  \Cref{theo:approx_3mKP}.}\label{eg:3mKP}
The tightness of guarantee \( \alpha\ge 1\) is trivial; any example where the algorithm gives an optimal solution to GMKP works.
Refer to \Cref{fig:mKP-eg0} for the tight \( \beta\le 1/3\) example.
\begin{figure}[ht]\caption{Tight \(\pmb{\beta\le 1/3}\) example for  \Cref{theo:approx_3mKP}}\label{fig:mKP-eg0}
	\begin{tikzpicture}[x=0.75pt,y=0.75pt,yscale=-1,xscale=1]
		
		\begin{scope}
			\node [anchor=west] at (70,65) {GMKP optimal solution};
			\draw (43,133) node   {$0$};
			\draw (43,83) node   {$1$};
			\draw    (60,81) -- (60,131) ;
			\draw    (60,81) -- (50,81) ;
			\draw    (60,131) -- (50,131) ;
			
			\draw (190.5,107) node   {$\cdots $};
			
			\draw  [color=gray] (80,81) -- (120,81) -- (120,131) -- (80,131) -- cycle ;
			\draw  [color=gray] (130,81) -- (170,81) -- (170,131) -- (130,131) -- cycle ;
			\draw  [color=gray] (210,81) -- (250,81) -- (250,131) -- (210,131) -- cycle ;
			\draw  [color=gray] (260,81) -- (300,81) -- (300,131) -- (260,131) -- cycle ;
		\end{scope}
		
		\begin{scope}
			\node [anchor=west] at (70,175) {\Cref{alg:3mKP_GMKP} bi-GMKP solution};	
			\draw (43,260) node   {$0$};
			\draw (43,210) node   {$1$};
			\draw (43,160) node   {$2$};		
			\draw    (60,160) -- (60,260) ;
			\draw    (60,160) -- (50,160) ;
			\draw    (60,210) -- (50,210) ;
			\draw    (60,260) -- (50,260) ;

			\draw  [color=gray] (80,210) -- (120,210) -- (120,260) -- (80,260) -- cycle ;
			\draw  [color=gray] (130,210) -- (170,210) -- (170,260) -- (130,260) -- cycle ;
			\draw  [color=gray] (210,210) -- (250,210) -- (250,260) -- (210,260) -- cycle ;			
			\draw  [fill=lightgray] (260,200) -- (300,200) -- (300,220) -- (260,220) -- cycle ;			
			\draw  [color=gray] (260,210) -- (300,210) -- (260,210) -- (300,210) -- cycle ;
			
			\draw  [fill=lightgray] (80,220) -- (120,220) -- (120,260) -- (80,260) -- cycle ;
			\draw  [fill=lightgray] (130,220) -- (170,220) -- (170,260) -- (130,260) -- cycle ;
			\draw  [fill=lightgray] (210,220) -- (250,220) -- (250,260) -- (210,260) -- cycle ;
			\draw  [fill=lightgray] (260,220) -- (300,220) -- (300,260) -- (260,260) -- cycle ;
			
			\draw (100,240) node [scale=1]  {$\frac{3m-1}{3m}$};
			\draw (150,240) node [scale=1]  {$\frac{3m-1}{3m}$};
			\draw (190.5,235) node   {$\cdots $};
			\draw (230,240) node [scale=1]  {$\frac{3m-1}{3m}$};
			\draw (280,240) node [scale=1]  {$\frac{3m-1}{3m}$};
			\draw (280,210) node [scale=.95]  {$\frac{1}{3}$};
			
			\node[scale=0.8] [anchor=west] at (299,205) {$\}$};
			\node [anchor=west] at (305,205) {$\frac{3m-1}{3m}+\frac{1}{3}-1=\frac{1}{3}-\frac{1}{3m}$};
		\end{scope}
		
		\begin{scope}[shift={(-105,0)}]%legend
			\draw (530,102) node  [align=left] {Items of group 1};
			\draw (500,142) node  [align=left] {Capacity};
			
			\draw  [color=gray] (440,130) -- (460,130) -- (460,150) -- (440,150) -- cycle ;
			\draw  [fill=lightgray] (440,90) -- (460,90) -- (460,110) -- (440,110) -- cycle ;
			
			\draw   (430,80) -- (600,80) -- (600,160) -- (430,160) -- cycle ;
		\end{scope}
		
	\end{tikzpicture}
	
\end{figure}
Consider \( m\ge 3\)  knapsacks of equal capacities 1,  and one group with \(m\) items that weigh \((3m-1)/({3m})\) and one item of weight \(1/3\).
\Cref{alg:3mKP_GMKP} generates a solution where one  knapsack has the item of weight \(1/3\) assigned and one item that weighs \(({3m-1})/({3m})\).
Therefore, the maximum exceeded knapsack capacity is \(1/3-1/({3m})\),  and as \( m\to \infty \) the bound converges to \( 1/3\).
Note constraints~\eqref{eq:KP-GMKP-cap2.2},~\eqref{eq:2mKP-GMKP-constr2}, and~\eqref{eq:3mKP-GMKP-constr} of 3mKP-GMKP are satisfied.

\paragraph{Examples for  \Cref{theo:approx_mKP-GMKP2}.}\label{eg:2mKPD}
Refer to \Cref{fig:mKPD-eg} for the tight \( \beta\le 1/2\) example.
\begin{figure}[ht]\caption{Tight \(\pmb{\beta\le 1/2}\) example for  \Cref{theo:approx_mKP-GMKP2}}\label{fig:mKPD-eg}
	
	\tikzset{every picture/.style={line width=0.75pt}} %set default line width to 0.75pt        
	
	\begin{tikzpicture}[x=0.75pt,y=0.75pt,yscale=-1,xscale=1]
		\begin{scope}
			\draw  [fill=lightgray] (460,50) -- (480,50) -- (480,70) -- (460,70) -- cycle ;
			\draw  [color=gray] (460,90) -- (480,90) -- (480,110) -- (460,110) -- cycle ;
			\draw   (450,40) -- (620,40) -- (620,120) -- (450,120) -- cycle ;
			\draw (550,62) node  [align=left] {Items of group 1};
			\draw (513,102) node  [align=left] {Capacity};
		\end{scope}
		
		\begin{scope}
			\node [anchor=west] at (60,45) {GMKP optimal solution};
			
			%0 to 1 lines
			\draw    (59.88,61) -- (59.88,111) ;
			\draw    (59.88,111) -- (49.88,111) ;
			\draw    (59.88,61) -- (49.88,61) ;
			\draw (42.88,111) node   {$0$};
			\draw (42.88,61) node   {$1$};
			
			% items
			\draw  [fill=lightgray] (69.88,61) -- (130,61) -- (130,86) -- (69.88,86) -- cycle ;
			\draw  [fill=lightgray] (69.88,86) -- (130,86) -- (130,111) -- (69.88,111) -- cycle ;
			\draw  [fill=lightgray] (140,65) -- (200,65) -- (200,110) -- (140,110) -- cycle ;
			\draw  [fill=lightgray] (210,70) -- (270,70) -- (270,110) -- (210,110) -- cycle ;
			\draw (287,95) node   {$\cdots $};
			\draw  [fill=lightgray] (300,75) -- (360,75) -- (360,110) -- (300,110) -- cycle ;
			\draw  [fill=lightgray] (370,81) -- (430,81) -- (430,110) -- (370,110) -- cycle ;
			
			%item sizes
			\draw (100,73) node   {$\frac{1}{2}  $};
			\draw (100,99) node   {$\frac{1}{2}  $};
			\draw (170,90) node   {$\frac{2m-1}{2m}  $};
			\draw (240,92) node   {$\frac{2m-2}{2m} $};
			\draw (330,94) node   {$\frac{m+2}{2m} $};
			\draw (400,96) node   {$\frac{m+1}{2m} $};
		\end{scope}
		
		\begin{scope}[shift={(0,-20)}]
			\node [anchor=west] at (60,150) {\Cref{alg:mKPD-GMKP} bi-GMKP solution};
			
			%0 to 1 lines 
			\draw    (60,165) -- (50,165) ;
			\draw    (60,165) -- (60,215) ;
			\draw    (60,215) -- (50,215) ;
			\draw (43,215) node   {$0$};
			\draw (42,165) node   {$1$};
			%Knapsacks
			
			%Shape: Rectangle [id:dp8790176395638236] 
			\draw  [color=gray] (70,165) -- (130.12,165) -- (130.12,215) -- (70,215) -- cycle ;
			\draw  [color=gray] (140.12,169) -- (200.12,169) -- (200.12,214) -- (140.12,214) -- cycle ;
			\draw  [color=gray] (210.12,174) -- (270.12,174) -- (270.12,214) -- (210.12,214) -- cycle ;
			\draw (287,199) node   {$\cdots $};
			\draw  [color=gray] (300.12,179) -- (360.12,179) -- (360.12,214) -- (300.12,214) -- cycle ;
			
			% items
			\draw  [fill=lightgray] (70,171) -- (130.12,171) -- (130.12,215) -- (70,215) -- cycle ;
			\draw  [fill=lightgray] (140.12,175) -- (200.24,175) -- (200.24,214) -- (140.12,214) -- cycle ;
			\draw  [fill=lightgray] (210,180) -- (270.12,180) -- (270.12,214) -- (210,214) -- cycle ;
			\draw  [fill=lightgray] (300,185) -- (360.12,185) -- (360.12,214) -- (300,214) -- cycle ;
			\draw  [fill=lightgray] (370.12,190) -- (430.24,190) -- (430.24,214) -- (370.12,214) -- cycle ;
			\draw  [fill=lightgray] (370.12,166) -- (430.24,166) -- (430.24,190) -- (370.12,190) -- cycle ;
			
			\draw   (434.75,184.72) .. controls (437.33,184.72) and (438.62,183.43) .. (438.62,180.85) -- (438.62,180.85) .. controls (438.62,177.16) and (439.91,175.32) .. (442.49,175.32) .. controls (439.91,175.32) and (438.62,173.48) .. (438.62,169.79)(438.62,171.45) -- (438.62,169.79) .. controls (438.62,167.21) and (437.33,165.92) .. (434.75,165.92) ;
			\draw (510,175) node   {$1-\frac{m+1}{2m}=
					\frac{1}{2} -\frac{1}{2m}  $};
			
			%last knapsack draws over items
			\draw  [color=gray] (370.12,185) -- (430.12,185) -- (430.12,185) -- (370.12,185) -- cycle ;

			%item sizes
			\draw (100,195) node   {$\frac{2m-1}{2m}  $};
			\draw (170,196) node   {$\frac{2m-2}{2m}$};
			\draw (240,198) node   {$\frac{2m-3}{2m} $};
			\draw (330,201) node   {$\frac{m+1}{2m} $};
			\draw (400,203) node   {$\frac{1}{2}  $};
			\draw (400,178) node   {$\frac{1}{2}  $};
			
		\end{scope}
	\end{tikzpicture}
	
\end{figure}

Consider \( m\ge 3\)   knapsacks were capacities are
\( c_i=({2m+1-i})/({2m}) \), \( \forall i\in M\), and
consider a single group with \( m+1\)  items, whose weights are
\( w_j=({2m-j})/({2m})= c_j-1/2m\), \( \forall j\in N\setminus \{n\} \)  and \( {w_n=1/2}\).
The group is feasible in GMKP instance, since the first \( m\)  items \( {j\in N\setminus \{n\}}\) can be assigned respectively to knapsacks \( j+1\)  where they fit exactly, and both items with \( w_{n-1}=w_n=1/2\) can be assigned to the first knapsack of size 1.
On the other hand, the algorithm sequentially assigns each item \( j\in N\setminus \{n\} \)   to knapsack \( i=j\).
Before assigning the last item \( n\), all knapsacks have \( 1/({2m})\)  free capacity, so assigning \( n\)  anywhere exceeds the capacity by \( 1/2 -1/({2m})\).
Having \( m\to\infty \)  gets bound \( 1/2\).
This example is also tight for the \(\alpha\ge 1\) bound.

\paragraph{Examples for  \Cref{theo:approx_mKP-GMKP3}.}\label{eg:3mKPD}
The tightness of guarantee \( \alpha\ge 1\) is trivial; any example where the algorithm gives an optimal solution to GMKP works.
\begin{figure}[ht]\caption{Tight \(\pmb{\beta\le 1/3}\) example for  \Cref{theo:approx_mKP-GMKP3}}\label{fig:mKP-GMKP_equal_tight_example}
	
	\tikzset{every picture/.style={line width=0.75pt}} %set default line width to 0.75pt        
	
	\begin{tikzpicture}[x=0.75pt,y=0.75pt,yscale=-1,xscale=1]
		\begin{scope}[shift={(-30,15)}]%legend
			\draw  [fill=lightgray] (460,50) -- (480,50) -- (480,70) -- (460,70) -- cycle ;
			\draw  [pattern=linevertical]  (460,90) -- (480,90) -- (480,110) -- (460,110) -- cycle ;
			\draw  [pattern=linedownleftright] (460,130) -- (480,130) -- (480,150) -- (460,150) -- cycle ;
			\draw  [color=gray] (460,170) -- (480,170) -- (480,190) -- (460,190) -- cycle ;
			\draw   (450,40) -- (620,40) -- (620,200) -- (450,200) -- cycle ;
			\draw (550,62) node  [align=left] {Items of group 1};
			\draw (550,102) node  [align=left] {Items of group 2};
			\draw (550,142) node  [align=left] {Items of group 3};
			\draw (513,182) node  [align=left] {Capacity};
		\end{scope}
		
		\begin{scope}
			\node [anchor=west] at (60,45) {GMKP optimal solution};
			
			\draw    (59.88,61) -- (59.88,111) ;
			\draw    (59.88,111) -- (49.88,111) ;
			\draw    (59.88,61) -- (49.88,61) ;
			\draw (42.88,111) node   {$0$};
			\draw (42.88,61) node   {$1$};
			
			\node [anchor=center] at (230,85) {(Depends on $D$. One group is removed)};
		\end{scope}
		\begin{scope}
			\node [anchor=west] at (60,130) {\Cref{alg:3mKP_GMKP} bi-GMKP solution};
			
			\draw    (60,165) -- (50,165) ;
			\draw    (60,165) -- (60,215) ;
			\draw    (60,215) -- (50,215) ;
			\draw	 (43,215) node   {$0$};
			\draw	 (42,165) node   {$1$};
			
			%Knapsacks
			\draw  [color=gray] (130,165) -- (180,165) --
			(180,214) -- (130,214) -- cycle ;
			\draw (200,190) node   {$\cdots $};
			\draw  [color=gray] (220,165) -- (270,165) --
			(270,214) -- (220,214) -- cycle ;
			\draw (290,190) node   {$\cdots $};
			\draw  [color=gray] (310,165) -- (360,165) --
			(360,214) -- (310,214) -- cycle ;
			\draw (380,190) node   {$\cdots $};
			
			% items
			\draw  [fill=lightgray] (70,171) -- (120,171) -- (120,215) -- (70,215) -- cycle ;
			\draw  [fill=lightgray] (70,148) -- (120,148) -- (120,171) -- (70,171) -- cycle ;
			\draw  [fill=lightgray] (130,171) -- (180,171) -- (180,214) -- (130,214) -- cycle ;

			\draw  [pattern=linevertical]  (220,165) -- (270,165) --
			(270,189.5) -- (220,189.5) -- cycle ;
			\draw  [pattern=linevertical] (220,189.5) -- (270,189.5) --
			(270,214) -- (220,214) -- cycle ;
			
			\draw  [pattern=linedownleftright]  (310,165) -- (360,165) -- 		(360,188) -- (310,188) -- cycle ;
			\draw  [pattern=linedownleftright] (310,188) -- (360,188) -- 		(360,214) -- (310,214) -- cycle ;

			\draw   (121.75,164.32) .. controls (123.94,164.33) and (125.05,163.23) .. (125.06,161.04) -- (125.06,161.04) .. controls (125.07,157.9) and (126.18,156.34) .. (128.37,156.35) .. controls (126.18,156.34) and (125.09,154.76) .. (125.1,151.63)(125.09,153.04) -- (125.1,151.63) .. controls (125.11,149.43) and (124.01,148.33) .. (121.82,148.32) ;
			
			\draw   (130.37,217.15) .. controls (130.39,221.82) and (132.73,224.14) .. (137.4,224.12) -- (160.07,224.02) .. controls (166.74,223.99) and (170.08,226.3) .. (170.1,230.97) .. controls (170.08,226.3) and (173.4,223.95) .. (180.07,223.92)(177.07,223.94) -- (202.73,223.82) .. controls (207.4,223.8) and (209.72,221.46) .. (209.7,216.79) ;
			\node [anchor=center] at (170,245) {$\times\left(\left\lceil\tfrac{1}{3\epsilon}\right\rceil-1\right)$};
			
			\draw   (220.03,216.82) .. controls (220.06,221.49) and (222.4,223.81) .. (227.07,223.78) -- (249.73,223.68) .. controls (256.4,223.65) and (259.74,225.97) .. (259.76,230.64) .. controls (259.74,225.97) and (263.06,223.62) .. (269.73,223.59)(266.73,223.6) -- (292.4,223.49) .. controls (297.07,223.47) and (299.39,221.13) .. (299.37,216.46) ;
			\node [anchor=center] at (260,245) {$\times\left\lfloor\tfrac{q_{\max}}{3}\right\rfloor$};
			
			\draw   (310.32,216.82) .. controls (310.34,221.49) and (312.68,223.81) .. (317.35,223.78) -- (340.02,223.68) .. controls (346.69,223.65) and (350.03,225.97) .. (350.05,230.64) .. controls (350.03,225.97) and (353.35,223.62) .. (360.02,223.59)(357.02,223.6) -- (382.68,223.49) .. controls (387.35,223.47) and (389.67,221.13) .. (389.65,216.46) ;
			\node [anchor=west] at (310,245) {$\times \left\lfloor\tfrac{2p+1}{3}\right\rfloor, \forall\tfrac{1}{2p+1}\in D_\text{odd}$};
			
			\node [anchor=west] at (127,155) {$\frac{1}{3}-\epsilon$};
			
			\draw  [color=gray] (70,165) -- (120,165) -- (120,165) -- (70,165) -- cycle ;
			
			\draw (95,159) node   {$\tfrac{1}{3}$};
			\draw (95,192) node   {$1-\epsilon$};
			
			\draw (155,192) node   {$1-\epsilon$};
			
			\node[scale=.9] at (245,177) {\contour{white}{$\tfrac{1}{2}$}};
			\node[scale=.9] at (245,202) {\contour{white}{$\tfrac{1}{2}$}};
			
			\node[scale=.9] at (335,177) {\contour{white}{$\tfrac{p}{2p+1}$}};
			\node[scale=0.9] at (335,202) {\contour{white}{$\tfrac{p+1}{2p+1}$}};
			
		\end{scope}
	\end{tikzpicture}
	
\end{figure}

Refer to \Cref{fig:mKP-GMKP_equal_tight_example} for the tight \( \beta\le 1/3\) example, where all knapsacks have equal capacities of 1.
Recall from \Cref{prop:minset_D} that only elements of \(D\) of the form \({c_{\max}/q=1/q}\), \(q\in \mathbb{Z}_{>0}\), are relevant to be considered.
Let \(D\) be partitioned into \(D_{\text{odd}}\) and \(D_{\text{even}}\),
where for all \( 1/q\in  D_{\text{odd}}\), \(q\) is odd; and for all \( 1/q\in  D_{\text{even}}\), \(q\) is even.
Let \(1/q_{\max}\in D_{\text{even}}\) be the smallest number in \( D_{\text{even}}\).

For a small \(\epsilon>0\), consider \(m = \left\lceil1/(3\epsilon) \right\rceil +\left\lfloor q_{\max}/3 \right\rfloor+\sum_{1/q\in D_\text{odd}} \left\lfloor q/3\right\rfloor\) knapsacks and groups where
\begin{itemize}
	\item Group 1 has an item that weighs \(1/3\) and \(\left\lceil 1/({3\epsilon}) \right\rceil \) items that weigh \( 1-\epsilon\).
	\item Group 2 has \(2\left\lfloor q_{\max}/3 \right\rfloor\) items that weigh \( 1/2\).
	\item
	      For each \(1/({2p+1})\in D_\text{odd}\),
	      group 3 has \( \left\lfloor({2p+1})/{3}\right\rfloor\) items that weigh \( ({p+1})/({2p+1})\), and \( \left\lfloor({2p+1})/{3}\right\rfloor\) items that weigh \( {p}/({2p+1})\).
\end{itemize}
If all rewards equal total group weights then,
given this ordering of groups,
\Cref{alg:2mKP-GMKP} generates a solution where all groups are selected.
One knapsack has an item of weight \( 1-\epsilon\)  and another of weight \( 1/3\).
Therefore, the maximum exceeded knapsack capacity is \( 1/3-\epsilon\), so as \( \epsilon\to 0 \) it converges to bound \( 1/3\).

We show the solution is feasible in the \( \text{mKP}_\text{D}\)-GMKP instance.
Capacity constraint~\eqref{eq:KP-GMKP-cap3} is satisfied
\begin{align*}
	\text{lhs}
	 & =\sum_{l\in K}\sum_{j\in G_l} w_j z_l
	\\
	 & =\tfrac{1}{3}
	+
	\left\lceil\tfrac{1}{3\epsilon} \right\rceil (1-\epsilon)
	+
	2\left\lfloor\tfrac{q_{\max}}{3} \right\rfloor \tfrac{1}{2}
	+
	\sum_{\frac{1}{2p+1}\in D_\text{odd}}
	\left\lfloor\tfrac{2p+1}{3}\right\rfloor
	\left(
	\tfrac{p+1}{2p+1}
	+
	\tfrac{p}{2p+1}
	\right)
	\\
	 & \le \left\lceil\tfrac{1}{3\epsilon} \right\rceil
	+
	\left\lfloor\tfrac{q_{\max}}{3} \right\rfloor
	+
	\sum_{\frac{1}{q}\in D_\text{odd}}
	\left\lfloor\tfrac{q}{3}\right\rfloor
	=m= \sum_{i\in M} c_i
	=rhs
\end{align*}
Constraints~\eqref{eq:mKPD-GMKP-constr} are satisfied for any \( 1/({2p})\in D_\text{even}\)
\begin{align*}
	 & \text{lhs} =
	\sum_{l\in K}\sum_{j\in G_l} f_\frac{1}{2p}(w_j) z_l
	\\
	 & =
	f_\frac{1}{2p}\left(\tfrac{1}{3}\right)
	+
	\left\lceil\tfrac{1}{3\epsilon} \right\rceil f_\frac{1}{2p}(1-\epsilon)
	+
	2\left\lfloor\tfrac{q_{\max}}{3} \right\rfloor f_\frac{1}{2p}\left(\tfrac{1}{2}\right)
	+
	\sum_{\frac{1}{2p'+1}\in D_\text{odd}}
	\left\lfloor\tfrac{2p'+1}{3}\right\rfloor
	\left(
	f_\frac{1}{2p}\left(\tfrac{p'+1}{2p'+1}\right)
	+
	f_\frac{1}{2p}\left(\tfrac{p'}{2p'+1}\right)
	\right)
	\\
	 & \le
	\left\lfloor\tfrac{2p}{3}\right\rfloor
	+
	\left\lceil\tfrac{1}{3\epsilon} \right\rceil (2p-1)
	+
	2\left\lfloor\tfrac{q_{\max}}{3} \right\rfloor (p-1)
	+
	\sum_{\frac{1}{q}\in D_\text{odd}}
	\left\lfloor\tfrac{q}{3}\right\rfloor
	f_\frac{1}{2p}(1)
	\\
	 & \le
	\left\lfloor\tfrac{q_{\max}}{3}\right\rfloor
	+
	\left\lceil\tfrac{1}{3\epsilon} \right\rceil (2p-1)
	+
	\left\lfloor\tfrac{q_{\max}}{3} \right\rfloor (2p-2)
	+
	\sum_{\frac{1}{q}\in D_\text{odd}}
	\left\lfloor\tfrac{q}{3}\right\rfloor
	(2p-1)
	\\
	 & =
	\left(
	\left\lceil\tfrac{1}{3\epsilon} \right\rceil
	+
	\left\lfloor\tfrac{q_{\max}}{3} \right\rfloor
	+
	\sum_{\frac{1}{q}\in D_\text{odd}}
	\left\lfloor\tfrac{q}{3}\right\rfloor
	\right)(2p-1)
	=
	m(2p-1)
	=
	\sum_{i\in M} f_\frac{1}{2p}(1)
	=
	\sum_{i\in M} f_\frac{1}{2p}(c_i)
	=
	\text{rhs}
\end{align*}
Constraints~\eqref{eq:mKPD-GMKP-constr} are satisfied for any \( 1/({2p+1})\in D_\text{odd}\)
\begin{align*}
	\text{lhs} = & 
	\sum_{l\in K}\sum_{j\in G_l} f_\frac{1}{2p+1}(w_j) z_l
	\\
	=            & 
	f_\frac{1}{2p+1}\left(\tfrac{1}{3}\right)
	+
	\left\lceil\tfrac{1}{3\epsilon} \right\rceil f_\frac{1}{2p+1}(1-\epsilon)
	+
	2\left\lfloor\tfrac{q_{\max}}{3} \right\rfloor f_\frac{1}{2p+1}\left(\tfrac{1}{2}\right)
	\\& +
	\sum_{\frac{1}{2p'+1}\in D_\text{odd}}
	\left\lfloor\tfrac{2p'+1}{3}\right\rfloor
	\left(
	f_\frac{1}{2p+1}\left(\tfrac{p'+1}{2p'+1}\right)
	+
	f_\frac{1}{2p+1}\left(\tfrac{p'}{2p'+1}\right)
	\right)
	\\
	\le          & 
	\left\lfloor\tfrac{2p+1}{3}\right\rfloor
	+
	\left\lceil\tfrac{1}{3\epsilon} \right\rceil 2p
	+
	2\left\lfloor\tfrac{q_{\max}}{3} \right\rfloor p
	\\& 
	+
	\left\lfloor\tfrac{2p+1}{3}\right\rfloor
	\left(
	f_\frac{1}{2p+1}\left(\tfrac{p+1}{2p+1}\right)
	+
	f_\frac{1}{2p+1}\left(\tfrac{p}{2p+1}\right)
	\right)
	+
	\sum_{\frac{1}{q}\in D_\text{odd}\setminus\left\{\frac{1}{2p+1}\right\}}
	\left\lfloor\tfrac{q}{3}\right\rfloor
	f_\frac{1}{2p+1}(1)
	\\
	\le          & 
	\left\lfloor\tfrac{2p+1}{3}\right\rfloor
	+
	\left\lceil\tfrac{1}{3\epsilon} \right\rceil 2p
	+
	2\left\lfloor\tfrac{q_{\max}}{3} \right\rfloor p
	+
	\left\lfloor\tfrac{2p+1}{3}\right\rfloor
	(2p-1)
	+
	\sum_{\frac{1}{q}\in D_\text{odd}\setminus\left\{\frac{1}{2p+1}\right\}}
	\left\lfloor\tfrac{q}{3}\right\rfloor
	2p
	\\
	=            & 
	\left(
	\left\lceil\tfrac{1}{3\epsilon} \right\rceil
	+
	\left\lfloor\tfrac{q_{\max}}{3} \right\rfloor
	+
	\sum_{\frac{1}{q}\in D_\text{odd}}
	\left\lfloor\tfrac{q}{3}\right\rfloor
	\right)
	2p
	=
	m2p
	=
	\sum_{i\in M} f_\frac{1}{2p+1}(1)
	=
	\sum_{i\in M} f_\frac{1}{2p+1}(c_i)
	=
	\text{rhs}
\end{align*}

\section{Results for Different Rewards}\label[appendix]{APPENDIX:results}

All instances solved in \Cref{SEC:exp_design} had rewards of each group equal to the total weight of items in that group.
Here we test the same instances after modifying each reward \(p_l, l\in K\), in three different ways:
\begin{itemize}
	\item Original Reward R0: \(p^0_l=\sum_{j\in G_l}w_j\)
	\item Reward R1: \(p^1_l=\lfloor 100\sqrt{p^0_l}\rceil\)
	\item Reward R2: \(p^2_l=\lfloor p^0_l\sqrt{p^0_l}\rceil\)
	\item Reward R3: \(p^3_l=\left\lfloor Random(1,10)\cdot p^0_l\right\rceil\)
\end{itemize}
Function \(\lfloor\cdot\rceil\) denotes rounding to the nearest integer, to avoid precision issues with the IP solver.
Groups with reward R1 have a reward to weight ratio of approximately \(100/\sqrt{p_l^0}\); giving an incentive to pick lighter groups (we multiplied by 100 to have more precision when rounding).
Groups with reward R2 have a reward to weight ratio of approximately \(\sqrt{p_l^0}\); giving an incentive to pick heavier groups.
Finally, reward R3 consists of multiplying the original rewards with a real random number between 1 and 10 (each \(p^3_l\) is multiplied by a different random number); adding noise to the instance while still having rewards proportional to weights in expectation.

We repeated all experiments from \Cref{SEC:exp_design} for an additional 9,000 instances, given by the combination of the original 3,000 instances and the three additional reward structures.
Only one instance for reward R1 and one for reward R2 were not solved by Gurobi due to memory limitations, so we removed them from the results.

\subsection{Results for bi-GMKP Algorithms}

\begin{figure}[H]\caption{Maximum exceeded knapsack capacity per bi-GMKP algorithm after swap-optimal improvement for reward R1}\label{fig:exceeded_cap0-R1}
	\includegraphics[width=\linewidth]{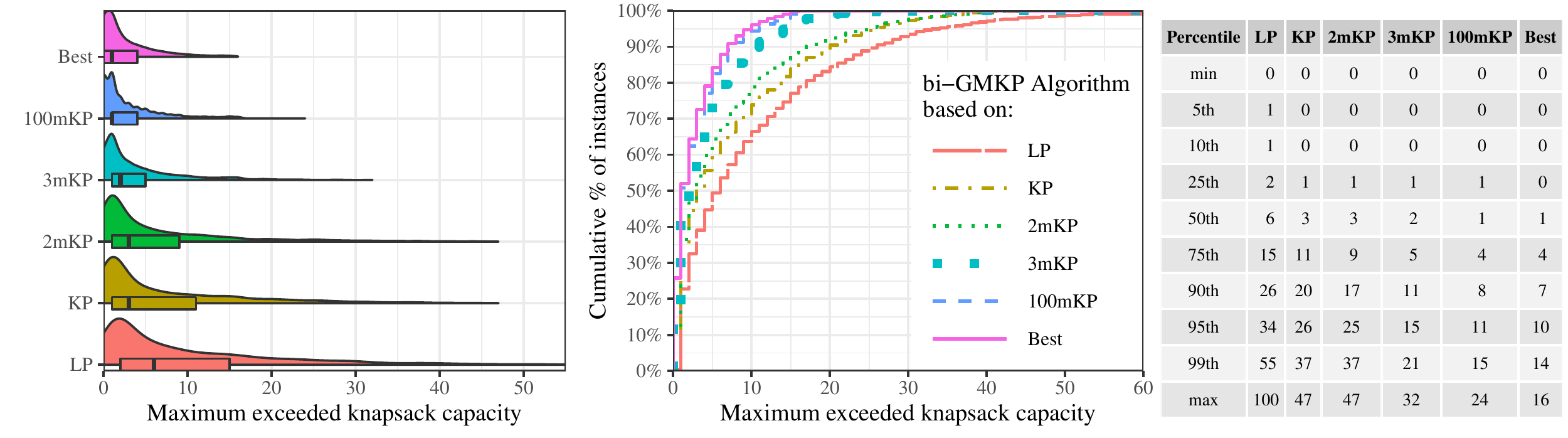}
\end{figure}

\begin{figure}[H]\caption{Maximum exceeded knapsack capacity per bi-GMKP algorithm after swap-optimal improvement for reward R2}\label{fig:exceeded_cap0-R2}
	\includegraphics[width=\linewidth]{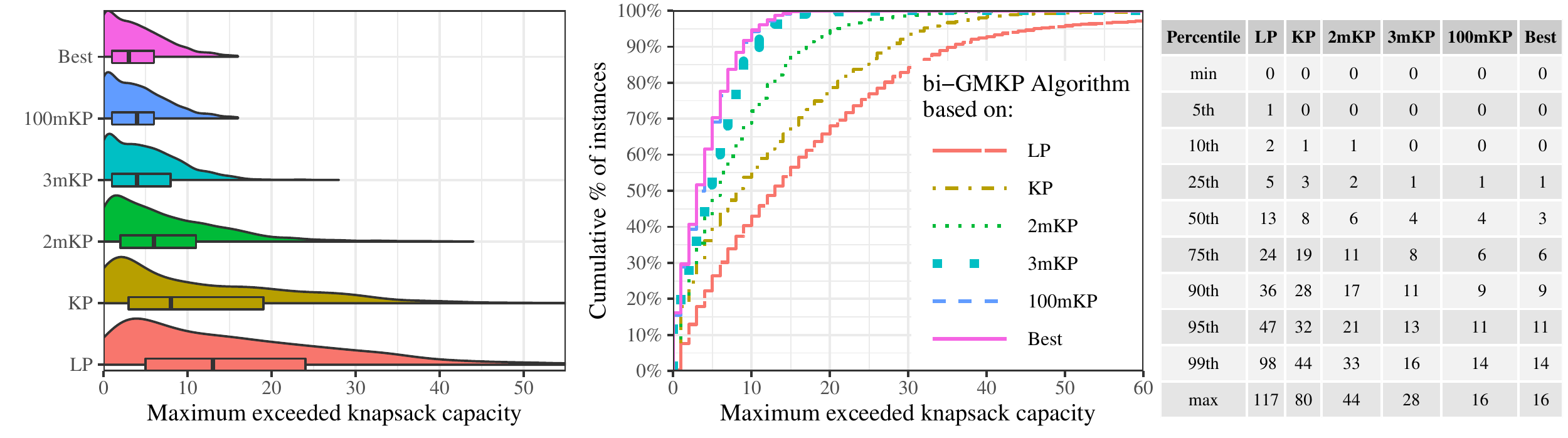}
\end{figure}

\begin{figure}[H]\caption{Maximum exceeded knapsack capacity per bi-GMKP algorithm after swap-optimal improvement for reward R3}\label{fig:exceeded_cap0-R3}
	\includegraphics[width=\linewidth]{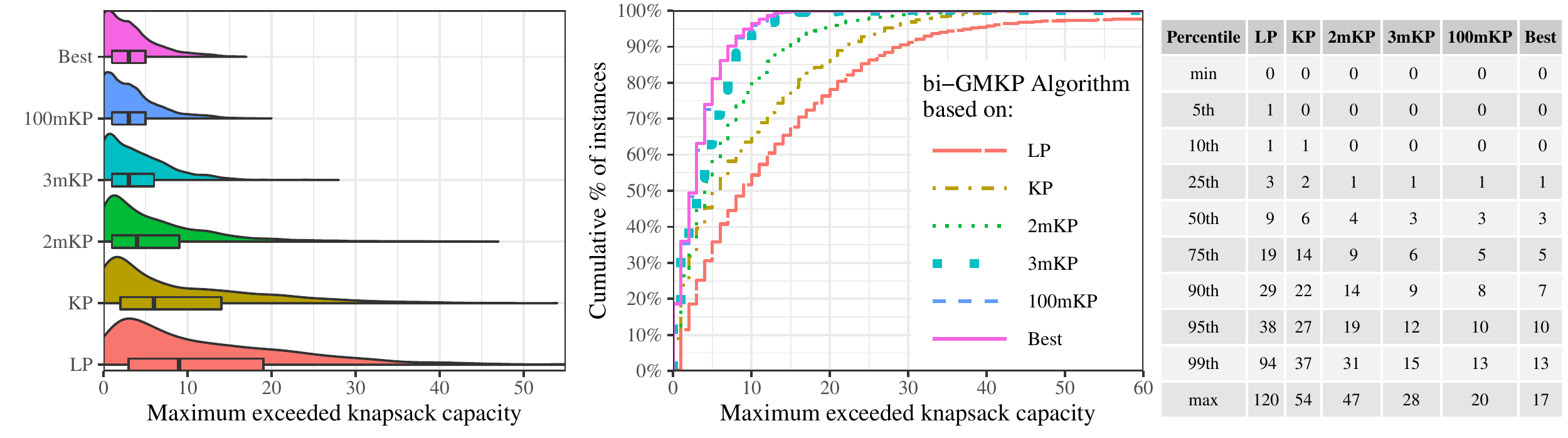}
\end{figure}

In \Cref{fig:exceeded_cap0-R1,fig:exceeded_cap0-R2,fig:exceeded_cap0-R3} we see the details of the maximum exceeded knapsack capacity of bi-GMKP algorithms, after swap-optimal improvement, for different reward structures.
Rewards R1 had solutions with lower maximum exceeded knapsack capacity than rewards R2, which makes sense since R1 prioritizes lighter groups while R2 prioritizes heavier groups.
Rewards R0 and R3 are somehow similar (see \Cref{fig:exceeded_cap2,fig:exceeded_cap0-R3}, respectively), showing that having the same reward to weight ratio in expectation seems to obtain similar results.
Algorithms 3mKP and 100mKP obtained the least exceeded knapsack capacity independent of reward structure.


\begin{figure}[H]\caption{Computation time per bi-GMKP algorithm after swap-optimal improvement for reward R1}\label{fig:comp_time-R1}
	\includegraphics[width=\linewidth]{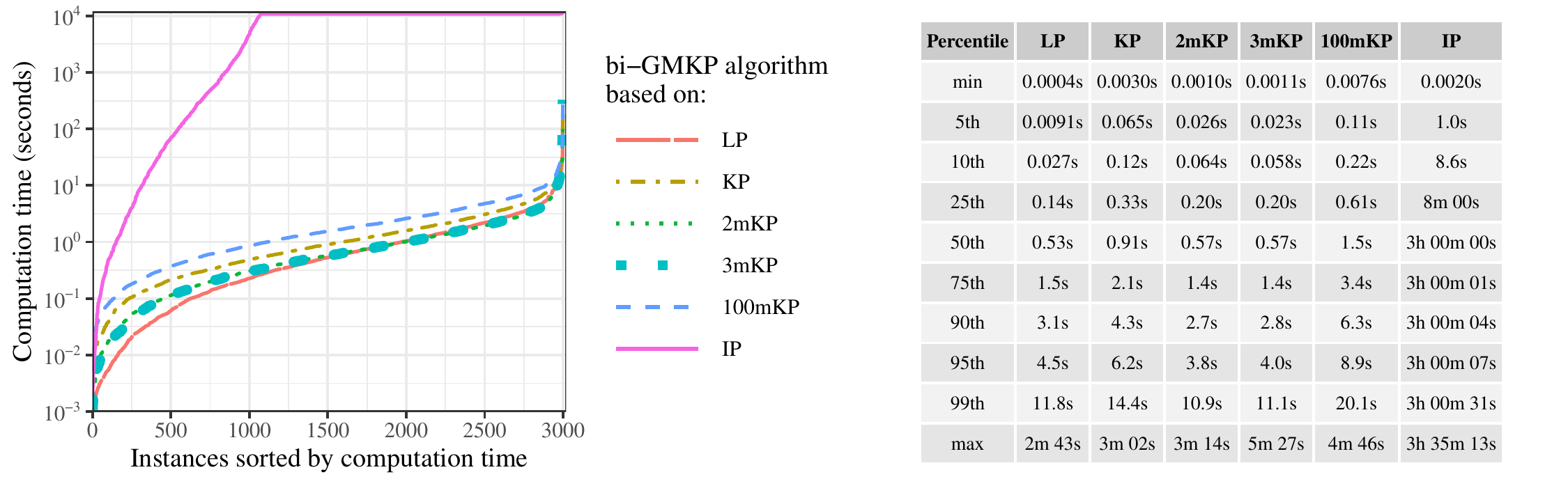}
\end{figure}

\begin{figure}[H]\caption{Computation time per bi-GMKP algorithm after swap-optimal improvement for reward R2}\label{fig:comp_time-R2}
	\includegraphics[width=\linewidth]{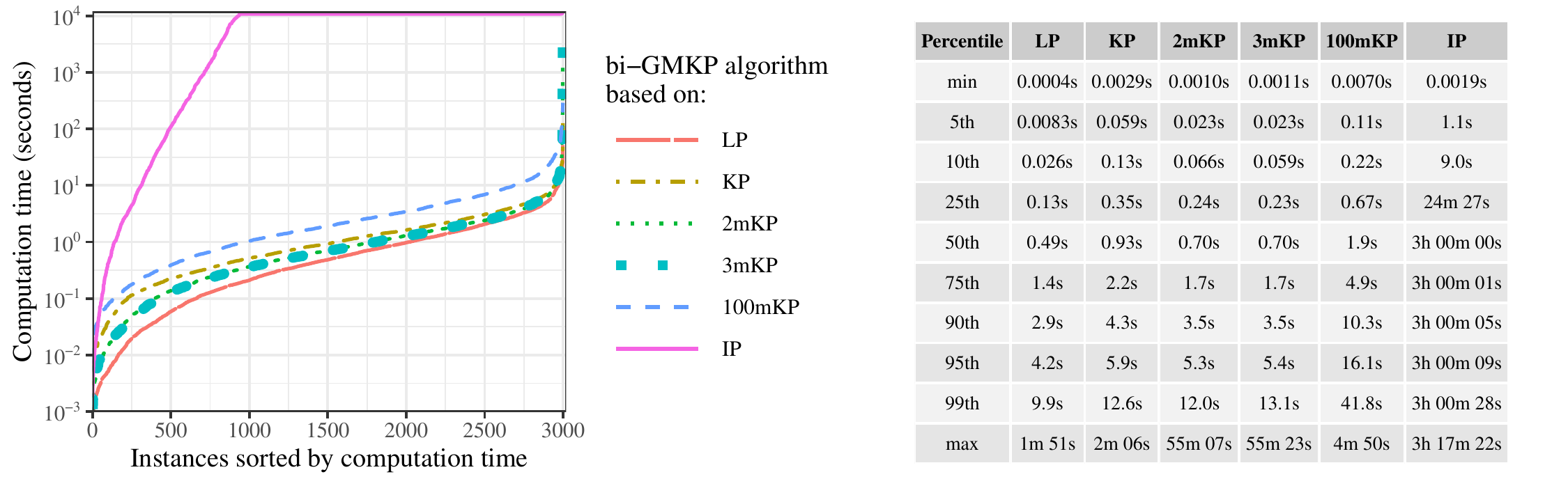}
\end{figure}

\begin{figure}[H]\caption{Computation time per bi-GMKP algorithm after swap-optimal improvement for reward R3}\label{fig:comp_time-R3}
	\includegraphics[width=\linewidth]{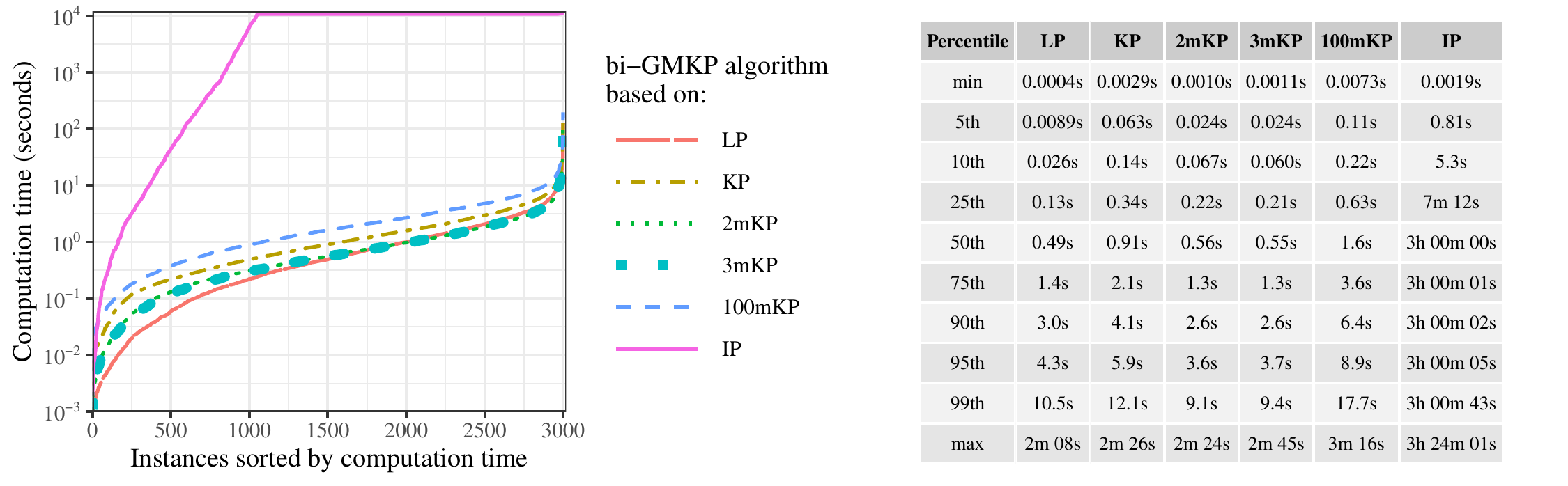}
\end{figure}

\Cref{fig:comp_time-R1,fig:comp_time-R2,fig:comp_time-R3} show the computation times of each bi-GMKP algorithm on instances with different reward structures.
Algorithms take a longer time solving instances where heavier groups are prioritized (reward R2), even obtaining outliers that take almost an hour to run; although 99\% of instances took under a minute.
3mKP persists as the most time-effective alternative independent of the reward structure, while running 100mKP might still be recommended since it obtains better results and computation times remain short.
It is interesting to note how for rewards R1, R2 and R3, Gurobi reached the time limit of 3 hours in around 67\% of instances, while in the original reward R0 (\Cref{fig:comp_time}) it only reached the time limit in about 33\%.
It seems that Gurobi works better when total group weights and rewards are equal.


\subsection{Results for GMKP Heuristics}
\begin{figure}[H]\caption{Optimal reward ratio per GMKP heuristic after swap-optimal improvement for reward R1}\label{fig:optimal_heur-R1}
	\includegraphics[width=\linewidth]{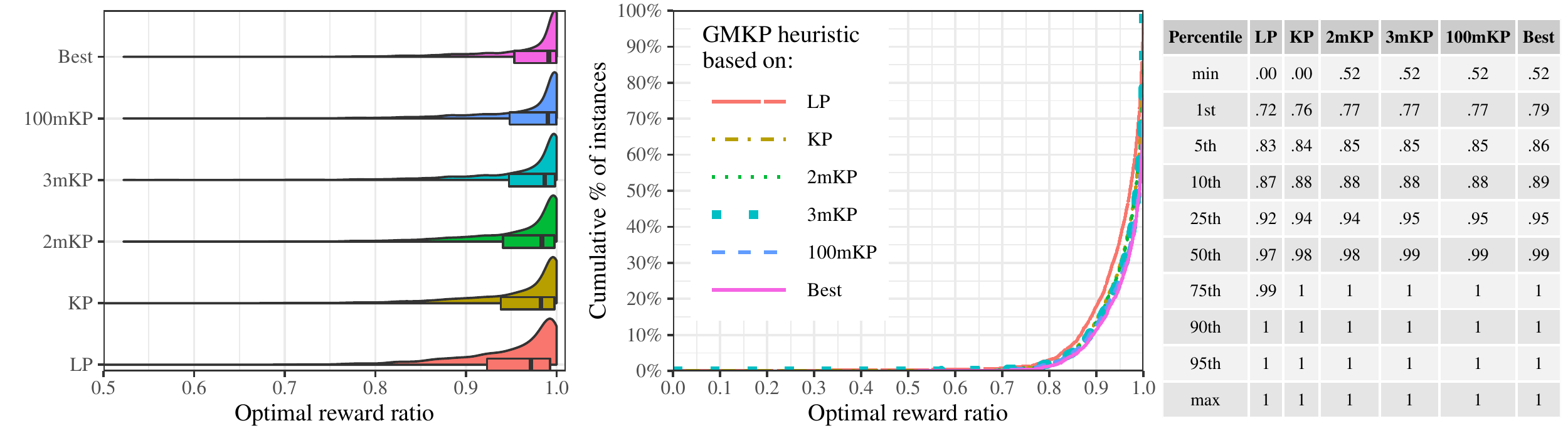}
\end{figure}

\begin{figure}[H]\caption{Optimal reward ratio per GMKP heuristic after swap-optimal improvement for reward R2}\label{fig:optimal_heur-R2}
	\includegraphics[width=\linewidth]{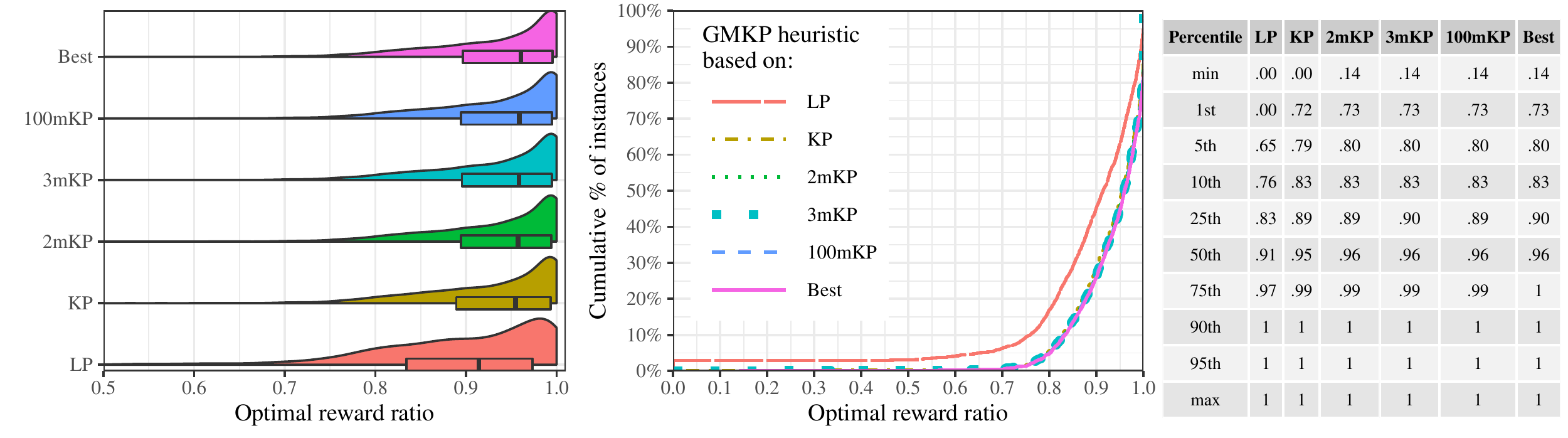}
\end{figure}

\begin{figure}[H]\caption{Optimal reward ratio per GMKP heuristic after swap-optimal improvement for reward R3}\label{fig:optimal_heur-R3}
	\includegraphics[width=\linewidth]{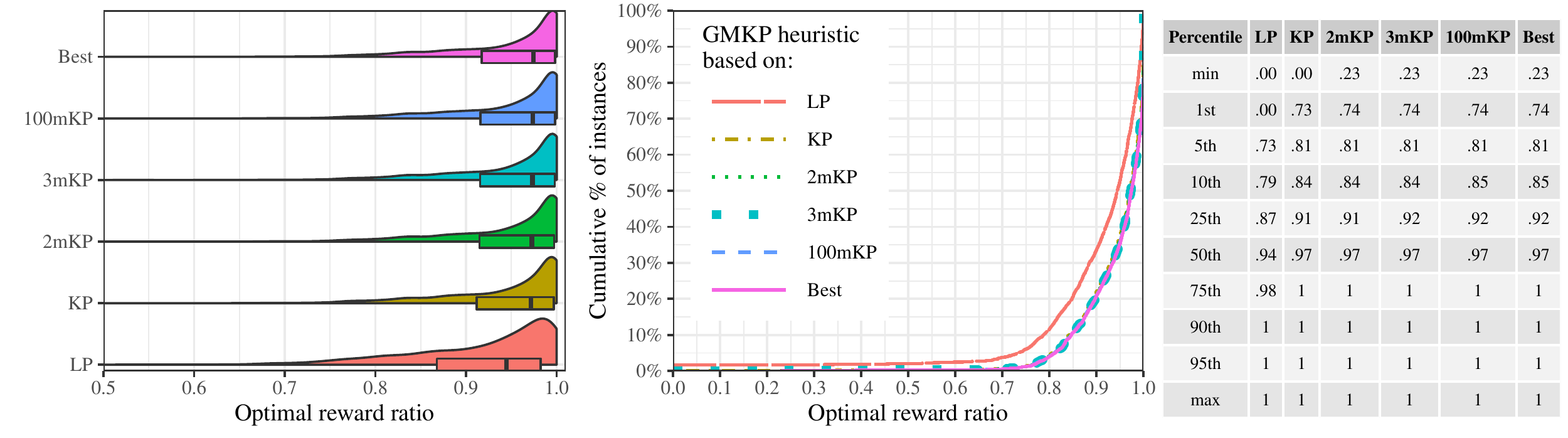}
\end{figure}

\Cref{fig:optimal_heur-R1,fig:optimal_heur-R2,fig:optimal_heur-R3} show the details of the optimal reward ratio of bi-GMKP heuristics, after swap-optimal improvement, for different reward structures.
As in the original reward R0 (\Cref{fig:optimal_heur}), there does not seem to be an improvement after adding constraints beyond 2mKP in any reward structure.
GMKP heuristics obtained the best performance when prioritizing smaller groups (R1), and random noise on rewards did not affect the performance significantly (R4).

The performance of the best GMKP heuristic dropped slightly in comparison to the original reward R0; 
the 5th percentile of optimal reward ratio was 0.83 in R0 (i.e., 95\% of instances did better than 0.83) while for rewards R1, R2, and R3 the 5th percentile dropped to 0.79, 0.73, and 0.74 respectively.
This difference might not be due to a loss in performance, but because most instances were not solved to optimality by the IP solver in rewards R1, R2, and R3 and their gap obtained was larger;
95\% of instances had a gap of 4.3\% or less in reward R0,
while the 95th percentile of gaps increased to 9.1\%, 11.1\%, and 8.2\% for rewards R1, R2, and R3 respectively.


\begin{figure}[H]\caption{Computation time per GMKP heuristic after swap-optimal improvement for reward R1}\label{fig:comp_time2-R1}
	\includegraphics[width=\linewidth]{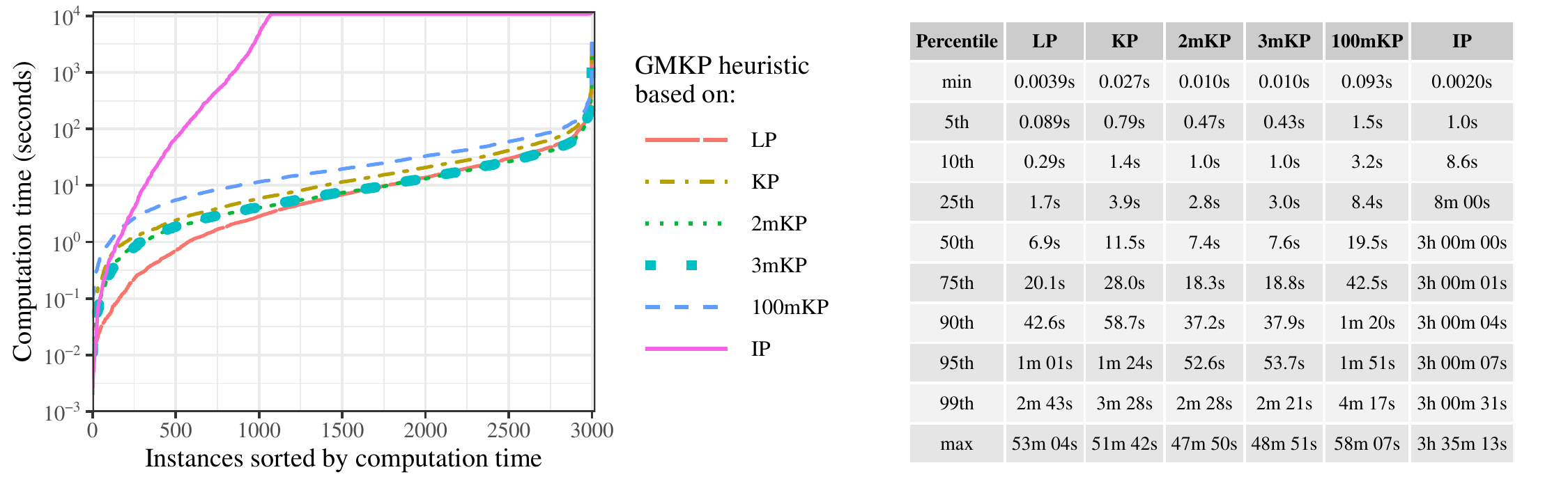}
\end{figure}

\begin{figure}[H]\caption{Computation time per GMKP heuristic after swap-optimal improvement for reward R2}\label{fig:comp_time2-R2}
	\includegraphics[width=\linewidth]{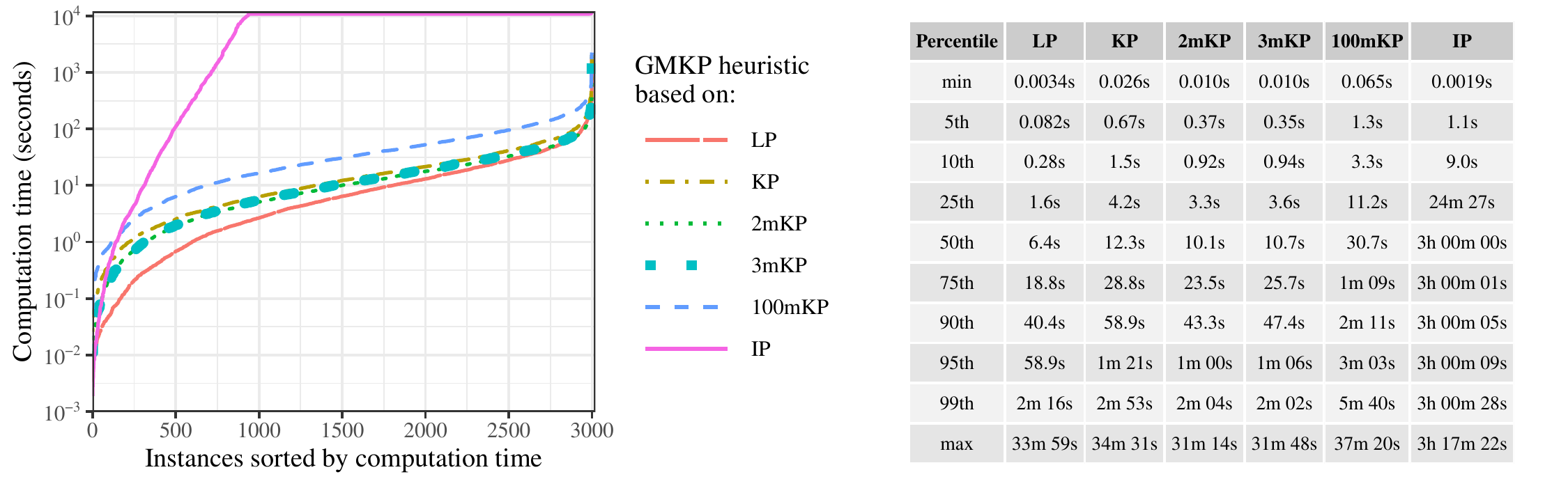}
\end{figure}

\begin{figure}[H]\caption{Computation time per GMKP heuristic after swap-optimal improvement for reward R3}\label{fig:comp_time2-R3}
	\includegraphics[width=\linewidth]{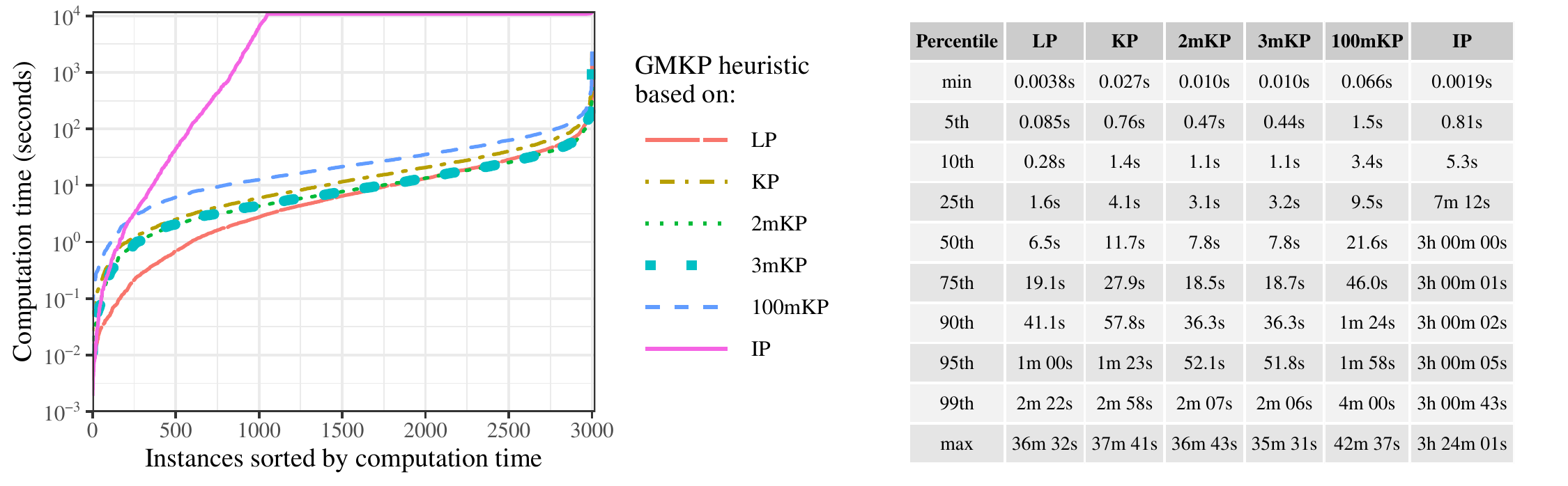}
\end{figure}

\Cref{fig:comp_time2-R1,fig:comp_time2-R2,fig:comp_time2-R3} show the computation times of each GMKP heuristic on instances with different reward structures.
2mKP runs faster than KP and similar to 3mKP, independent of reward structure.
Computation times increased in comparison to the original reward structure (see \Cref{fig:comp_time2}), where 95\% of instances were solved in less than 103 seconds in reward R0, and in less than 111, 183, and 118 seconds in reward R1, R2, and R3 respectively.
This difference might also be explained by Gurobi having better performance when total group weights equal rewards (recall sub-problems are also solved with Gurobi).


\subsection{Results for bi-GMKP Heuristics}

\begin{figure}[H]\caption{Non-dominated solutions of all instances  for reward R1}\label{fig:Pareto_example-R1}
	\includegraphics[width=0.48\textwidth]{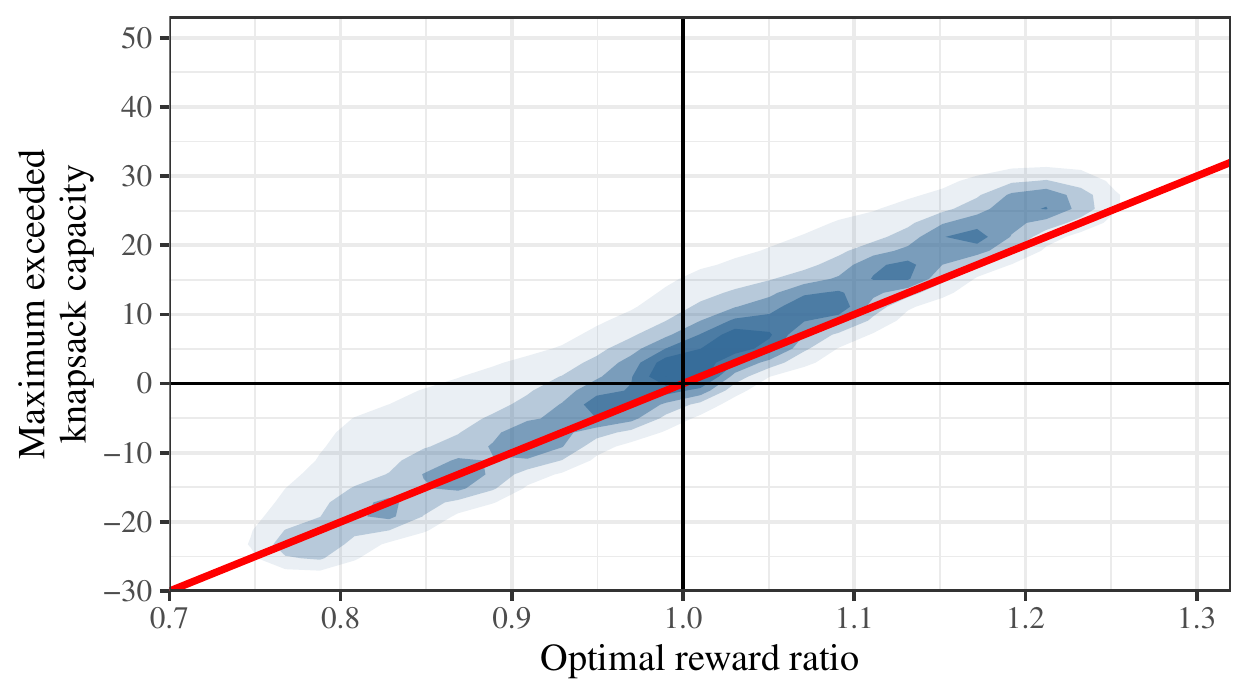}
\end{figure}

\begin{figure}[H]\caption{Non-dominated solutions of all instances  for reward R2}\label{fig:Pareto_example-R2}
	\includegraphics[width=0.48\textwidth]{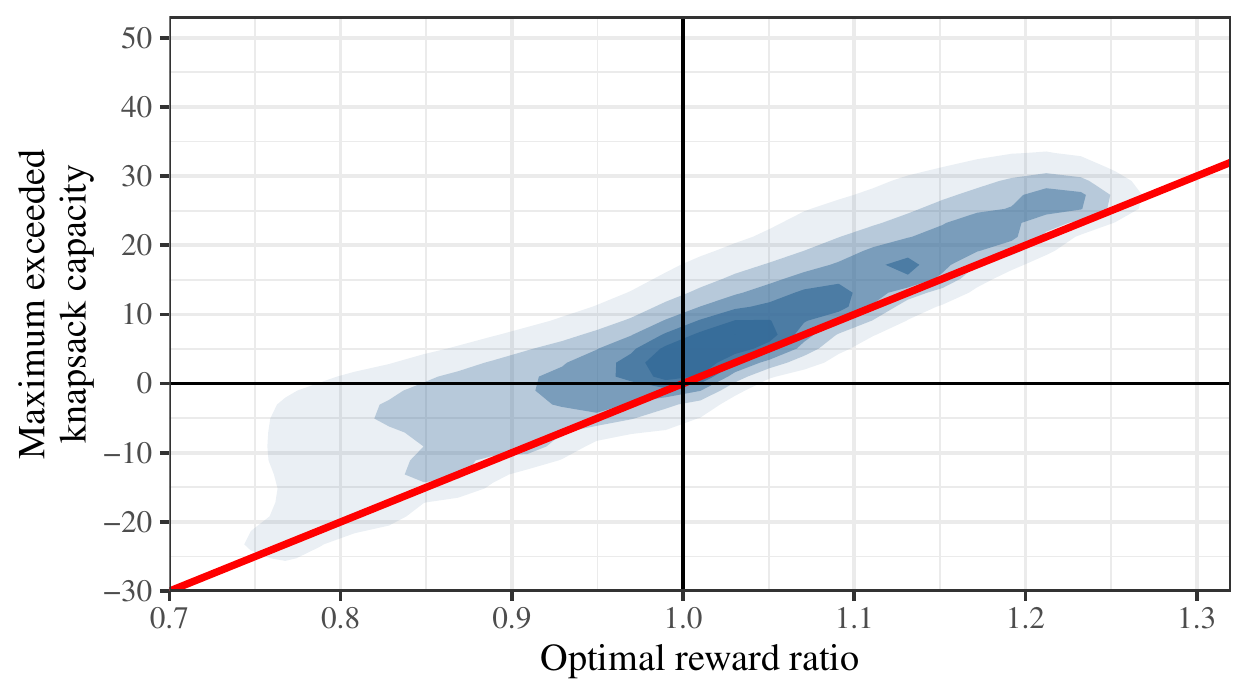}
\end{figure}

\begin{figure}[H]\caption{Non-dominated solutions of all instances  for reward R3}\label{fig:Pareto_example-R3}
	\includegraphics[width=0.48\textwidth]{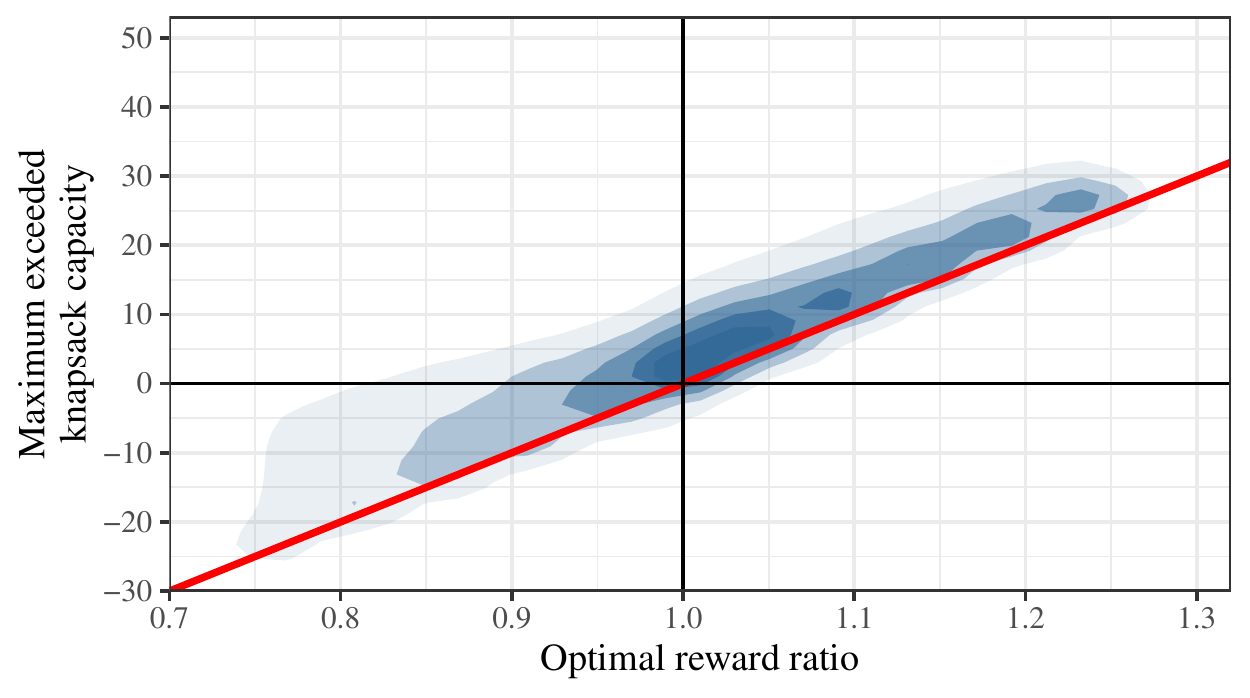}
\end{figure}

For different reward structures, \Cref{fig:Pareto_example-R1,fig:Pareto_example-R2,fig:Pareto_example-R3}
show the density of the non-dominated solutions obtained by bi-GMKP heuristic for the modified versions of \Cref{alg:2mKP-GMKP} (2mKP); as seen in \cref{alg-line:binary-GMKP} of \hyperref[heur:binary-GMKP2]{Heuristic~\ref*{heur:binary-GMKP}}.
Analogous to reward R0 (see \Cref{fig:Pareto_example}), most solutions lie slightly above the red line that represents the case where changes in the optimal reward ratio generate a proportional change in the maximum exceeded knapsack capcity.
This shows how the proposed bi-GMKP heuristic can be used, independent of the reward structure, to generate different bi-criteria combinations doing a good job in maximizing rewards while slightly exceeding knapsack capacities.


\end{appendices}

\end{document}